\documentclass[aps,
prd,
amssymb,
amsmath,
amsfonts,
eqsecnum,
showpacs,
nofootinbib,
floatfix,
twocolumn,
twoside,
a4paper,
superscriptaddress,
longbibliography
]{revtex4-1}

\usepackage[T3,T1]{fontenc}
\usepackage{mathrsfs} \usepackage{bm} \makeatletter
\@ifclassloaded{beamer}
  { \typeout{UsePackages: Detected beamer}
    \usepackage{tgheros}
    
  }
  { \typeout{UsePackages: Did not detect beamer}
    \ifx\asybeamer\undefined \typeout{UsePackages: Detected article}
    \usepackage[varg]{txfonts} \usepackage{tgtermes} \else \typeout{Fonts: Detected asy for beamer}
    \usepackage{tgheros}
    
    \fi
  }
\usepackage{microtype} \def\MT@register@subst@font{
  \MT@exp@one@n\MT@in@clist\font@name\MT@font@list
  \ifMT@inlist@\else\xdef\MT@font@list{\MT@font@list\font@name,}\fi}
\makeatother

\usepackage{graphicx} \usepackage[x11names,svgnames,rgb]{xcolor} \usepackage{xspace}
\usepackage{braket}
\usepackage{accents}
\usepackage{siunitx}
\usepackage{mathtools}
\DeclareSymbolFontAlphabet{\mathrm}{operators}

\definecolor{CiteColor}{rgb}{0.18039, 0.18824, 0.57255}
\definecolor{UrlColor} {rgb}{0.741, 0.173, 0.000}
\definecolor{DarkUrlColor} {rgb}{0.500, 0.110, 0.000}
\definecolor{LinkColor}{rgb}{0.25098, 0.47843, 0.04706}

\makeatletter \newcommand{\ShowFont}{\typeout{The main font is \f@encoding \space \f@family \space \f@series \space \f@shape \space at \f@size pt.}\typeout{The math font sizes are \tf@size pt (main), \sf@size pt (script), and \ssf@size pt (scriptscript).}\typeout{The linewidth is \the\linewidth}} \makeatother

\usepackage{xargs}

 \usepackage{graphicx}\usepackage{dcolumn}\usepackage{bm}\usepackage{tabularx}
\usepackage{multirow}
\usepackage{capt-of}
\usepackage{color}
\usepackage{url}
\usepackage[utf8]{inputenc}
\usepackage{placeins}
\usepackage{soul}

\usepackage[nolist,nohyperlinks]{acronym}
\usepackage{xspace}
\usepackage[colorlinks=true, linkcolor=blue, urlcolor=blue, citecolor=blue]{hyperref}

\usepackage{subcaption}
\usepackage[export]{adjustbox}

\usepackage{orcidlink}

\graphicspath{{figures/}}

\DeclareMathAlphabet{\mathbfsf}{\encodingdefault}{\sfdefault}{bx}{sl}

\newcommand{\fsampling}{f_{\mathrm{samp}}}

\newcommand{\be}{\begin{equation}}
\newcommand{\ee}{\end{equation}}
\newcommand{\bea}{\begin{eqnarray}}
\newcommand{\eea}{\end{eqnarray}}

\newcommand{\fmin}{f_{\mathrm{min}}}
\newcommand{\fmax}{f_{\mathrm{max}}}

\newcommand{\hosc}{h_{\mathrm{osc}}}
\newcommand{\hstep}{h_{\mathrm{step}}}
\newcommand{\htm}{h_{\mathrm{tm}}}

\newcommand{\lambdas}{\lambda_{\mathrm{s}}}
\newcommand{\lambdaosc}{\lambda_{\mathrm{o}}}

\newcommand{\As}{A_{\mathrm{s}}}
\newcommand{\ts}{t_{\mathrm{s}}}
\newcommand{\sigmas}{\sigma_{\mathrm{s}}}

\newcommand{\Aosc}{A_{\mathrm{o}}}
\newcommand{\tosc}{t_{\mathrm{o}}}
\newcommand{\sigmaosc}{\sigma_{\mathrm{o}}}

\definecolor{light-gray}{gray}{0.95}

\definecolor{dodgerblue}{HTML}{1E90FF}
\definecolor{viennared}{HTML}{DA0A14}
\definecolor{ctorange}{HTML}{FF6C0C}
\definecolor{granadagreen}{HTML}{078931}
\definecolor{wales}{HTML}{ff0038}
\definecolor{valenciacfred}{HTML}{ee3524}
\definecolor{barcelonafcgold}{HTML}{edbb00}
\definecolor{jam}{HTML}{A50B5E}
\definecolor{austriawien}{HTML}{441678}

\AtBeginDocument{\hypersetup{
    citecolor=blue,
    linkcolor=blue,   
    urlcolor=blue}}

\newcommand{\UIB}{Departament de F\'isica, Universitat de les Illes Balears, IAC3, Carretera Valldemossa km 7.5, E-07122 Palma, Spain}
\newcommand{\ICE}
{Institut de Ci\`encies de l'Espai (ICE, CSIC), Campus UAB, Carrer de Can Magrans s/n, 08193 Cerdanyola del Vall\`es, Spain}
\newcommand{\milan}{Dipartimento di Fisica ``G. Occhialini'', 
Universit\`a degli Studi di Milano-Bicocca, Piazza della Scienza 3, 20126 Milano, Italy}
\newcommand{\infn}{INFN, Sezione di Milano-Bicocca, 
Piazza della Scienza 3, 20126 Milano, Italy}

\allowdisplaybreaks[1]

\begin{document}

\title[ML]{Mind the step: On the frequency-domain analysis of gravitational-wave memory waveforms}

\author{Jorge Valencia\,\orcidlink{0000-0003-2648-9759}}
\email{jorge.valencia@uib.es}
\affiliation{\UIB}

\author{Rodrigo Tenorio\,\orcidlink{0000-0002-3582-2587}}
\affiliation{\milan}
\affiliation{\infn}
\affiliation{\UIB}

\author{Maria Rossell\'o-Sastre\,\orcidlink{0000-0002-3341-3480}}
\affiliation{\UIB}

\author{Sascha Husa\,\orcidlink{0000-0002-0445-1971}}
\affiliation{\ICE}
\affiliation{\UIB}

\date{\today}

\begin{abstract}

Gravitational-wave memory is characterized by a signal component that persists after a transient signal has decayed. Treating such signals in the frequency domain is non-trivial, since discrete Fourier transforms assume periodic signals on finite time intervals. In order to reduce artifacts in the Fourier transform, it is common to use recipes that involve windowing and padding with constant values. Here we discuss how to regularize the Fourier transform in a straightforward way by splitting the signal into a given sigmoid function that can be Fourier transformed in closed form, and a residual which does depend on the details of the gravitational-wave signal and has to be Fourier transformed numerically, but does not contain a persistent component. 
We provide a detailed discussion of how to map between continuous and discrete Fourier transforms of signals that contain a persistent component.
We apply this approach to discuss 
the frequency-domain phenomenology of the $(\ell=2, m=0)$ spherical harmonic mode, which contains both a memory and an oscillatory ringdown component.
\end{abstract}

\maketitle

\section{Introduction}
\label{sec:Introduction}

\begin{figure}[h!]
    \centering
    \includegraphics[width=\columnwidth]{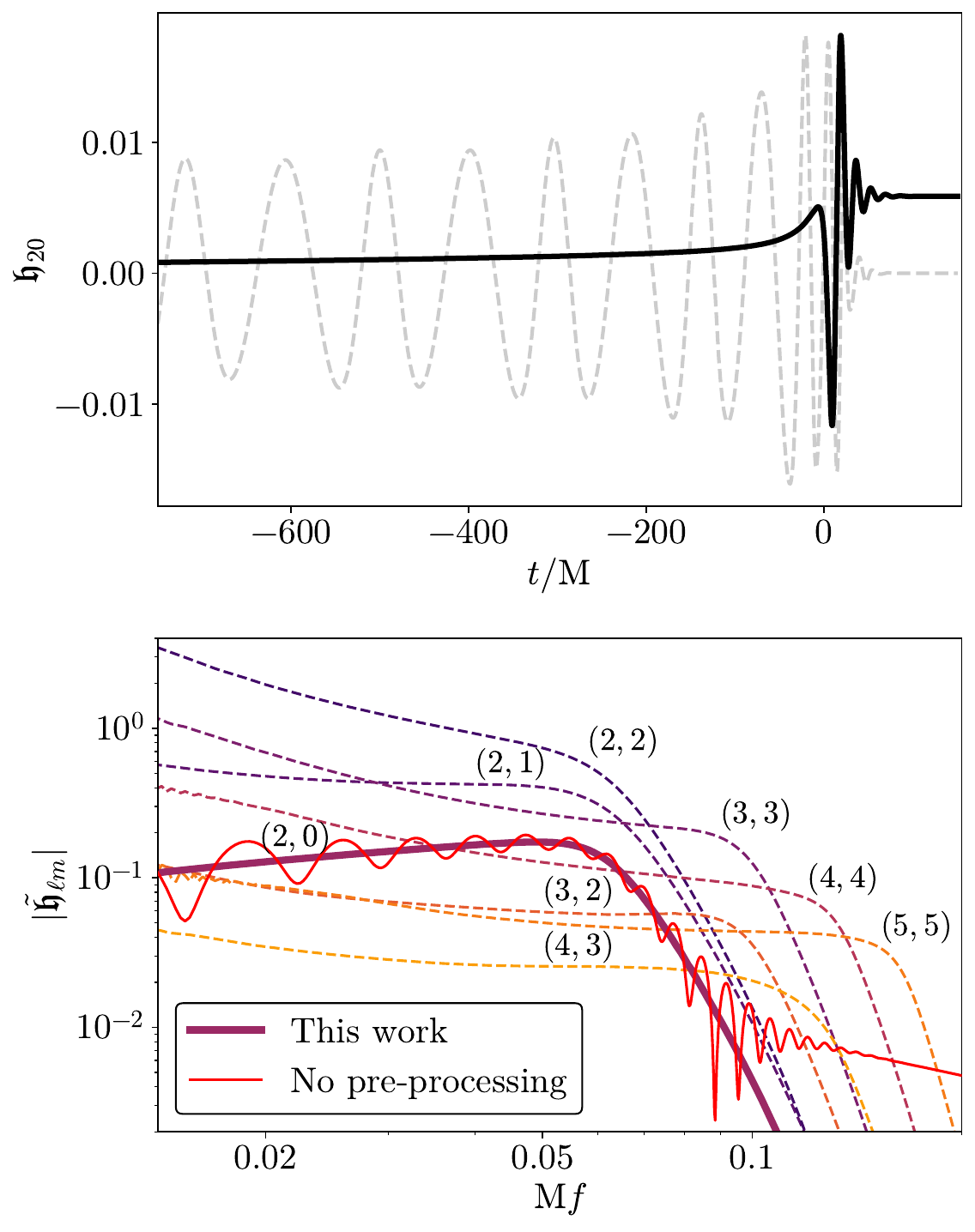}
    \caption{
        Spherical-harmonic decomposition of a GW signal
        from an aligned spin system with mass ratio 1:4
        and $\chi_1=\chi_2=-0.8$ (SXS:BBH:1936).
        The upper panel highlights the (2,0) mode in time domain
        with the (rescaled) non-persistent contributions separated in gray;
        the lower panel displays its Fourier transform together
        with the other modes. 
        The thick solid line is the frequency-domain $(2,0)$ mode after 
        applying the pre-processing presented in this work;
        the thin solid line lacks any pre-processing and displays a variety
        of artifacts due to the finite duration of the signal.
    }
    \label{fig:show_off}
\end{figure}

All gravitational-wave signals observed to date are transient signals believed to originate in compact binary coalescences \cite{LIGOScientific:2018mvr,Abbott:2020niy,LIGOScientific:2021usb,KAGRA:2021vkt}. General relativity however predicts that such signals also contain ``gravitational-wave memory'', a comparatively small component which persists after the transient has passed. Such signals thus exhibit a step-like behavior in the time domain, see for example the upper panel of Fig.~\ref{fig:show_off}.

Gravitational-wave memory is expected to be first observed in the next few years \cite{Lasky:2016knh,Hubner:2019sly,Boersma:2020gxx,Goncharov:2023woe,Grant:2022bla}, and it has created great interest due to the very different character of the signal as compared with the transient component, the nonlinear nature of the effect in the merger of bound objects \cite{Blanchet:1992br,PhysRevLett.67.1486}, and the connection with the Bondi-Metzner-Sachs (BMS) group of symmetries of asymptotically flat spacetimes.

Some of us have recently developed a computationally efficient phenomenological model of the $(\ell=2, m =0)$ spherical harmonic of quasicircular aligned spin 
coalescences of black holes \cite{Rossello-Sastre:2024zlr}. This mode contains the leading contribution of the gravitational-wave memory effect, as well as an 
oscillatory signal associated with quasi-normal ringdown, and completes the modeling of the $(\ell=2)$ modes. The model is constructed in the time domain;  in 
GW data analysis it is however very common to work in the frequency domain, e.g. a quantity of central interest is the following scalar product~\cite{Finn:1992wt}
\begin{equation}\label{eq:scalar_product}
    \langle x, y \rangle = 4\mathrm{Re}\int_{\fmin}^{\fmax} \mathrm{d} f \frac{\tilde{x}^{*}(f) \tilde{y}(f)}{S_{\mathrm{n}}(f)} \,,
\end{equation}
where $x$ and $y$ represent two arbitrary time series, and $S_{\mathrm{n}}(f)$ is the single-sided power spectral density. The latter effectively sets limits on the sensitive frequency range of a detector, and thus determines which types of sources can be observed by detectors such as Advanced LIGO~\cite{LIGOScientific:2014pky},
Advanced Virgo~\cite{VIRGO:2014yos}, KAGRA~\cite{KAGRA:2018plz}, ET~\cite{ET}, CE~\cite{CE}, 
or LISA~\cite{LISA}. In practice, the upper cutoff frequency $\fmax$ is often set by the signal, while the lower
cutoff $\fmin$ is a result of the detector's technology.

The importance of frequency-domain representations in gravitational-wave data analysis calls for an adequately simple and computationally efficient way to perform discrete Fourier transforms (FT) for step-like functions. A standard approach to carry out FTs that involve memory signals is to develop a recipe (see e.g.~\cite{Gasparotto_2023, Richardson_2022, chen2024improved}) based on windowing or padding with constant values to mitigate numerical artifacts. 
Following this approach, an ``optimal'' FT would need a laborious iterative experimentation process of choosing the appropriate window parameters and padding length tailored to the specific system, which creates a potential source of error.

Here we take a different route and focus on constructing the frequency-domain version of an infinitely long signal. 
In order to avoid FT artifacts we simply split the signal into a sigmoid function, which can be Fourier transformed in closed form and can be chosen a priori, and a residual which does depend on the details of the gravitational-wave signal and has to be Fourier transformed numerically, but does not contain the persistent  ``memory'' component. 
We then justify and provide cogent evidence for the success of our method using closed-form step-like signals and realistic
GW waveforms. Note that the response of space-borne detectors such as LISA~\cite{LISA},
which are based on time-delay interferometry~\cite{Tinto:2004wu}, does not generate step-like data for gravitational-wave memory signals, which allows alternative avenues to process those signals, see e.g.~\cite{Inchauspe:2024ibs}.

As an illustration of what our method achieves, the lower panel of Fig.~\ref{fig:show_off} shows the result of a 
straightforward numerical FT of a signal containing memory, along with the same result obtained 
using our method, which seamlessly suppresses numerical artifacts arising from the finite duration of the signal.
To understand how our approach relates to the common recipes based on windowing, we proceed by discussing a set of examples, where for each example the behavior of the numerical FT can be reproduced in closed form. In order to facilitate the use of our method we provide a Python implementation~\cite{FouTStep}.

This paper is organized as follows: 
We first briefly review basic properties of the FT in Sec.~\ref{sec:practicalities_ft}.
In Sec.~\ref{sec:toy_model} we apply our method to a simple toy model, where FTs can be carried out in closed form. This allows us to ensure that numerical FTs 
reproduce the analytical results. The gravitational memory case is then treated in Sec.~\ref{sec:memory}, which focuses on the FT of numerical relativity
(NR) waveforms. We summarize and discuss our results in Sec.~\ref{sec:discussion}.
Finally, in Appendices~\ref{app:shift} and~\ref{app:windows} we discuss further details of discrete FTs.

\section{Fourier transforms}\label{sec:practicalities_ft}

We define the FT to be consistent with the conventions adopted in the 
LIGO Algorithms Library~\cite{lalsuite}
\begin{equation}\label{eq:defFT}
    \tilde{x}(f) = \int_{-\infty}^{\infty} \mathrm{d}t \, x(t) \, e^{-i \, 2 \pi f t}\,.
\end{equation}
For any $k$-times differentiable function $x(t)$ its FT falloff can be characterized by a power law as ~\cite{TrefethenWeb1996}
\begin{equation}\label{eq:frequency_decays}
    \tilde{x}(f) = \mathcal{O}\left(f^{-(k+1)}\right) \,, \left| f \right| \rightarrow \infty \,
\end{equation}
while smooth functions will fall off faster than any polynomial.
As we will see below, in practice we will often encounter power-law falloffs,
e.g. due to boundary effects when working with FTs on finite domains.

The FT of time-shifted functions $x(t-t_0)$ gain an additional, frequency-dependent phase in the Fourier domain
with respect to Eq.~\eqref{eq:defFT}: 
\begin{equation}\label{eq:FT_shift_property}
    \int_{-\infty}^{\infty} \mathrm{d}t \, x(t - t_0) \, e^{-i \, 2 \pi f t} 
= e^{-i \, 2 \pi f t_0} \tilde{x}(f) \,.
\end{equation}

\subsection{Fourier transforms of step-like functions\label{subsec:closed_from}}

In order to prepare our treatment of the FT of step-like functions such as the memory signal shown in Fig.~\ref{fig:show_off}, 
we consider some examples where the FT can be carried out in closed form.
We start with the FT of the constant function $C$, given by
\begin{equation}
    \tilde{C}(f) = C \int_{-\infty}^{\infty} \mathrm{d}t\, e^{-i \, 2 \pi f t} = C \delta(f) \,.
\end{equation}
An overall offset of a function thus corresponds to an $f=0$ component
of the FT, usually known as the DC component. For our purposes, these effects
will always be negligible, as Eq.~\eqref{eq:scalar_product} is only affected
by frequencies $ f > \fmin > 0$.

Second, the FT of a Heaviside step function $H(t)$ reads
\begin{equation}\label{eq:FT_step}
    \tilde{H}(f)
    =  \int_{0}^{\infty} \mathrm{d}t \, e^{-i \, 2\pi f t}
    = \frac{1}{2} \delta(f) + \frac{1}{i \, 2 \pi f} 
\end{equation}
Once more we can neglect the delta and are left with a $\mathcal{O}(f^{-1})$ decay,
which arises from the fact that $H(t)$ is a discontinuous function and Eq.~(\ref{eq:frequency_decays}).

We can construct a smooth step by using a hyperbolic tangent
\begin{equation}
    s(t;\sigma) = \frac{1}{2} + \frac{1}{2}\mathrm{tanh}\left(\frac{t}{\sigma}\right) \,,
    \label{eq:smooth_step}
\end{equation}
where $\sigma > 0$ controls the timescale of the function's jump.
This is a regulator for the discontinuity, since this converges 
to $H(t)$ as $\sigma \rightarrow 0$. The corresponding FT is
\begin{equation}\label{eq:sigmoid_FT}
    \begin{aligned}
        \tilde{s}(f;\sigma) & = \frac{1}{2} \delta(f) - \frac{i \pi \sigma}{2} \, \mathrm{csch}(\sigma \pi^2 f) \\
        & =  \frac{1}{2} \delta(f) - \frac{i \pi \sigma}{e^{\pi^2 \sigma f} - e^{-\pi^2 \sigma f}} \,.
    \end{aligned}
\end{equation}
As expected from Eq.~(\ref{eq:frequency_decays}) the high frequency behavior \mbox{$\sigma f \gg 1$} is 
now dominated by an exponential decay
\begin{equation}
     \left. \tilde{s}(f; \sigma) \right|_{\sigma f \gg 1} \simeq -i \, \pi \sigma e^{-\pi^2 \sigma f}\,.
\end{equation}
To leading order, the low  frequency behavior \mbox{$\sigma f \ll 1$} becomes
\begin{equation}
    \left. \tilde{s}(f; \sigma) \right|_{\sigma f \ll 1} = \frac{1}{2} \delta(f) + \frac{1}{i \, 2\pi f} \left[1 + \mathcal{O}\left( (\sigma f)^2 \right) \right]\,.
\end{equation}
which coincides with Eq.~\eqref{eq:FT_step} as  $\sigma \rightarrow 0$.
Note the decay's amplitude is independent of the step's timescale, 
although the typical frequency up to which this behavior dominates
\emph{does} depend on $\sigma$ through $f \lesssim \sigma^{-1}$.
In short, step-like features primarily affect the low-frequency Fourier components;
the behavior at high frequencies, on the other hand, is governed by the smoothness of the function. 

With these results we can construct a simple window function for $t \in [t_0, t_0 + T]$
\begin{equation}
    w(t; t_0, T,\sigma) = s(t-t_0; \sigma) -  s(t - t_0 - T ;\sigma)\,.
    \label{eq:td_window} 
\end{equation}
The FT follows from Eqs.~\eqref{eq:FT_shift_property} and \eqref{eq:sigmoid_FT} in closed form:
\begin{equation}\label{eq:FTwindow}
    \tilde{w}(f; t_0, T, \sigma) = 
    e^{-i \, 2 \pi f t_0} \left(1 - e^{-i \,2 \pi f T} \right)\tilde{s}(f; \sigma) \,. 
\end{equation}
This closed-form derivation simplifies the task of analyzing 
the consequences of specific choices for the parameters.
Since we have combined two steps separated by a certain duration
and placed them at a certain initial time, Eq.~\eqref{eq:FTwindow} 
contains \emph{two} oscillatory factors. This shows that the FTs of step-like
signals with similar timescales \emph{will} interfere with each other.
In other words, window functions naturally introduce oscillatory artifacts in the frequency-domain signal. 
This will be especially relevant in Sec.~\ref{sec:toy_model} whenever GW memory is 
analyzed using a window, as then the expected $\mathcal{O}(f^{-1})$ 
falloff will gain a modulation as shown in Eq.~\eqref{eq:FTwindow}.

\subsection{The discrete Fourier Transform}\label{subsec:dft}

We are interested in observational data, which will be discretely sampled 
at a certain frequency $\fsampling$. This causes all Fourier amplitudes at frequencies separated by a multiple of $\fsampling / 2$ to fold onto each other. This phenomenon is usually known as frequency \emph{aliasing}, and does not pose a problem insofar data is band-limited within the detector's capabilities.

We represent a generic GW with a time series \mbox{$\{x_j=x(t_j), j=0, \dots, N-1\}$}
where each sample is labeled by a timestamp
$t_j = t_0 + j \delta t$.  $t_0$ is a fiducial start time and the number of samples is given by $N = (\delta t \fsampling)^{-1}$. The FT of this discrete dataset can then be computed as
\begin{align}\label{eq:defFiniteFT}
    \tilde{x}(f_k) 
    &= \int_{t_0}^{t_0 + T} \mathrm{d}t \, x(t) \, e^{-i \, 2 \pi f_k t}\\
    &=  e^{-i \, 2 \pi f_k t_0} \int_{0}^{T} \mathrm{d}t' \, x(t' + t_0) \, e^{-i \, 2 \pi f_k t'}\label{eq:shift}\\
    &\approx e^{-i\, 2\pi f_k t_0} \delta t \sum_{j=0}^{N-1} x_j e^{-i \, 2 \pi \frac{j k}{N}} \label{eq:dft}\\
    &= e^{-i\, 2\pi f_k t_0} \tilde{x}_k
    \label{eq:finite_ft}
\end{align}
where in Eq.~\eqref{eq:dft} the discretisation of the continuous integral results 
in the \emph{discrete} FT (DFT)
\begin{equation}\label{eq:defDFT}
    \tilde{x}_k= \delta t \sum_{j=0}^{N-1} x_j \, e^{-i \, 2 \pi \frac{j k}{N}} \,,
\end{equation}
which can be computed efficiently using FFT algorithms \cite{FFT}.

As opposed to  our original definition in Eq.~(\ref{eq:defFT})
the integral in Eq.~(\ref{eq:defFiniteFT}) is not symmetric; 
likewise, after substituting the continuous time variable by a discrete index,
the index starts at 0 and an oscillatory factor corresponding to a timeshift of
$t_0$ appears as a consequence of Eq.~(\ref{eq:FT_shift_property}).
This  oscillatory behavior may become inconvenient for practical 
applications and can be dealt with in closed form, as discussed in Appendix~\ref{app:shift}.

This correspondence between the FT and the DFT holds as long as the signals we deal
with are well contained within the observing time. See Sec.~\ref{subsec:splitting}
for how to do it when this is not the case.

\subsection{Example}\label{subsec:gaussian_example}

\begin{figure}
    \centering
    \includegraphics[width=\columnwidth]{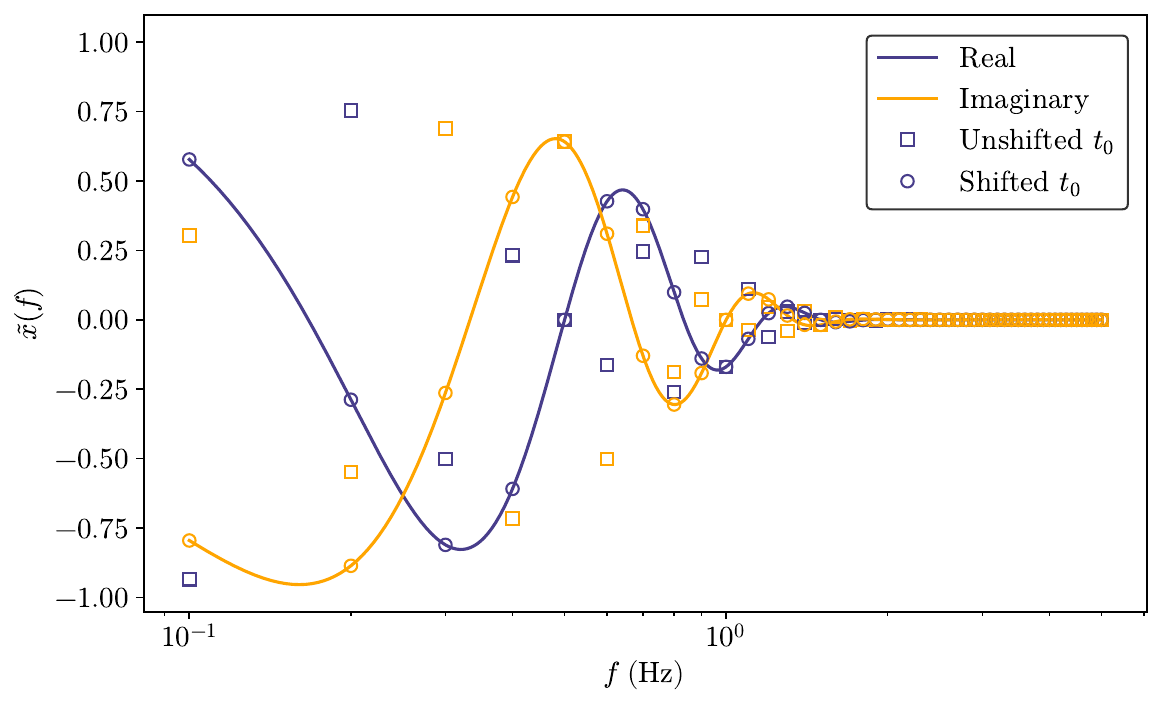}
    \caption{
        Comparison of discrete and continuous FT for a Gaussian function with $\mu=1.5$
        and $\sigma=0.3$. The time domain grid starts at $t_0 = -\SI{4}{\second}$ and samples
        $N = 100$ samples with a timestep of $\delta t = \SI{0.1}{\second}$. DFTs are evaluated
        at multiples of the Rayleigh frequency $1/(\delta t N)$.
        Since the input signal is real, only positive frequencies are shown.
        The solid lines
        show the continuous FT [Eq.~\eqref{eq:ft_gaussian}]. Square markers correspond to
        the DFT without accounting for the shifted time origin. Circle markers show the DFT
        including the phase terms steming from $t_0 \neq 0$.
    }
    \label{fig:discrete_gaussian}
\end{figure}

To illustrate these results we use the FT of a time-shifted Gaussian function and match the 
continuum and discrete results. We define our Gaussian function as 
\begin{equation}
    g(t; \mu, \sigma) = 
    \frac{1}{\sqrt{2 \pi \sigma^2}} e^{-\frac{1}{2}\left(\frac{t - \mu}{\sigma}\right)^2} \,,
    \label{eq:gauss}
\end{equation}
which has a corresponding continuous FT given by
\begin{equation}
    \tilde{g}(f;\mu,\sigma) = e^{-i\, 2 \pi f \mu} e^{-2 \pi^2 \sigma^2 f^2}
    \label{eq:ft_gaussian}
\end{equation}
Note that the presence of a non-zero mean $\mu$, which shifts the function away from the origin, maps directly into a frequency oscillation as shown in Eq.~\eqref{eq:FT_shift_property} and discussed above.

To compute the DFT, we evaluate Eq.~\eqref{eq:gauss} on a discrete grid with $N=100$ points
\mbox{$t_j = t_0 + j \delta t$} with fiducial values \mbox{$t_0 = -4$ and $\delta t = 0.1$}. 
The results are shown in Fig.~\ref{fig:discrete_gaussian}, from which we highlight two significant features.

First, while the FT of a Gaussian on a time-symmetric interval is again real and a Gaussian, our time-shifted version, where the start and end times are not symmetric with respect to $t=0$, picks up a complex oscillation, and it is not a real function.

Second, we note the excellent agreement between the discrete (and finite) FT and the continuous Fourier transform. As previously argued, this is due to the fact that the function has died out by the time we reach the end of the domain, and thus the truncation error is identically zero at the evaluated frequencies.
 
\section{How to analyze GW memory\label{sec:toy_model}}

Before treating gravitational-wave memory signals in Sec.~\ref{sec:memory},
we study a simple toy model of the (2,0)  mode in the time domain and with a closed-form FT.
This provides a ground truth to compare different pre-processing methods.
We first discuss \textit{windowing} (Sec.~\ref{subsec:windowing}),
which has some drawbacks when dealing with step-like functions such as GW memory.
To overcome those limitations we introduce Symbolic Sigmoid Subtraction \texttt{SySS} (Sec.~\ref{subsec:splitting}), 
which deals with the step symbolically in the time domain so artifacts are reduced. 
This makes it very convenient for GW memory waveforms. 
The code of this algorithm is publicly available as a Python package called \texttt{FouTStep} \cite{FouTStep}.

\subsection{A toy model for GW memory}

The morphology of the (2,0) mode in the time domain consists of a step-like behavior
($\hstep$) that leads into an oscillatory component ($\hosc$) at late times.
These two components are described in our toy model as 
\begin{equation}\label{eq:toy_td_step}
    \hstep(t;\As, \sigmas , \ts) 
    = \frac{\As}{2}\left[1 + \tanh\left(\frac{t-\ts}{\sigmas}\right)\right],
\end{equation}
and
\begin{equation}\label{eq:toy_td_osc}
    \hosc(t; \Aosc, \tosc, \sigmaosc, f_{\mathrm{o}}) 
    = \Aosc \sin[2\pi f_\mathrm{o} (t-\tosc)] e^{-\frac{1}{2}\left(\frac{t-\tosc}{\sigmaosc}\right)^2} \,,
\end{equation}
where $A_{\mathrm{s},\mathrm{o}}$ are the amplitude coefficients of the step and oscillating
components, $t_{\mathrm{s},\mathrm{o}}$ refer to their typical starting times,
$\sigma_{\mathrm{s}, \mathrm{o}}$ account for their typical timescales, and $f_\mathrm{o}$ refers
to the oscillation's frequency. We refer collectively to these two sets of parameters as 
$\lambdas$ and $\lambdaosc$, respectively. 
In the time domain the toy model then simply becomes
\begin{equation}\label{eq:toy_model}
    \htm(t;\lambdas, \lambdaosc) = \hstep(t;\lambdas) + \hosc(t;\lambdaosc)\,.
\end{equation}

\begin{figure}
    \centering
    \includegraphics[width=\columnwidth]{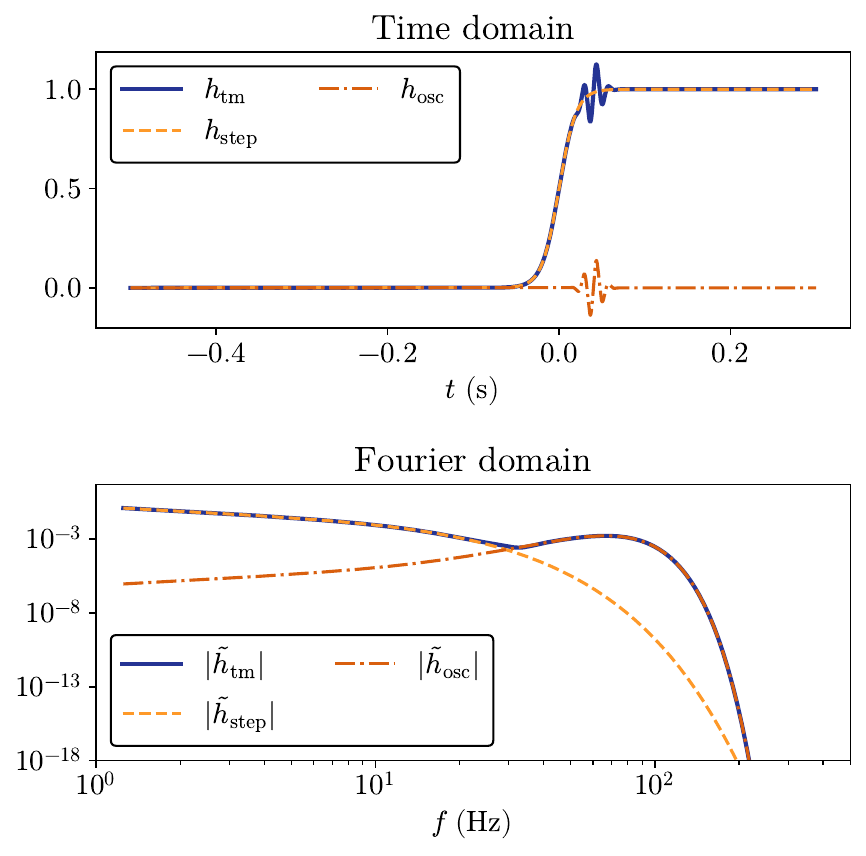}
    \caption{
        Representation of the GW memory toy model ($\htm$) together with its step ($\hstep$) and oscillatory ($\hosc$) components
        in both the time (upper panel) and frequency domain (lower panel). We have used the set of parameters 
        $\As=1,\, \ts=0\, \mathrm{s},\, \sigmas=0.02\, \mathrm{s}$, and
        $\Aosc=0.15,\, \tosc=0.04\, \mathrm{s},\, \sigmaosc=0.0177\, \mathrm{s},\, f_{\mathrm{o}}=66.7\, \mathrm{Hz}$.
        The time domain grid starts at $t_0=-0.5\, \mathrm{s}$ and samples $N=10^6$ samples with a timestep 
        of $\delta t=8\times10^{-6}\, \mathrm{s}$. The finite-length FTs $\tilde{h}_\mathrm{step}$ and $\tilde{h}_\mathrm{osc}$ 
        have been computed evaluating Eqs.~\eqref{eq:toy_fd_step} and~\eqref{eq:toy_fd_osc} at multiples of the Rayleigh frequency
        $1/(\delta t N)$. Since the time-domain input signals are real, we only show positive frequencies.
    }
    \label{fig:explain_toy_model}
\end{figure}

The FT of both components can be expressed in closed form using the results from 
Sec.~\ref{sec:practicalities_ft}:
\begin{align}
    \tilde{h}_{\mathrm{step}} (f;\lambdas) 
    & = -i \pi \sigmas \frac{\As}{2} \mathrm{csch}(\sigmas f \pi^2)\, e^{-i\, 2\pi f \ts}\,,
    \label{eq:toy_fd_step} \\
    \tilde{h}_{\mathrm{osc}}(f;\lambdaosc)
    & = -i \sqrt{2\pi}\sigmaosc \Aosc  e^{-i \, 2 \pi f \tosc - 2(\pi \sigmaosc)^2 (f^2 + f_{\mathrm{o}}^2)}
    \sinh(4\pi^2 \sigmaosc^2 f_{\mathrm{o}} f)  \label{eq:toy_fd_osc} \,.
\end{align}
We show an example of both the time and frequency-domain behavior of the toy model in Fig.~\ref{fig:explain_toy_model}. The
FT displays consistent behaviors with Sec.~\ref{sec:practicalities_ft}:
The step-like contribution dominates at low frequencies, and
the oscillating contribution dominates at high frequencies.

\begin{figure*}[p]
    \includegraphics[width=0.955\textwidth, right]{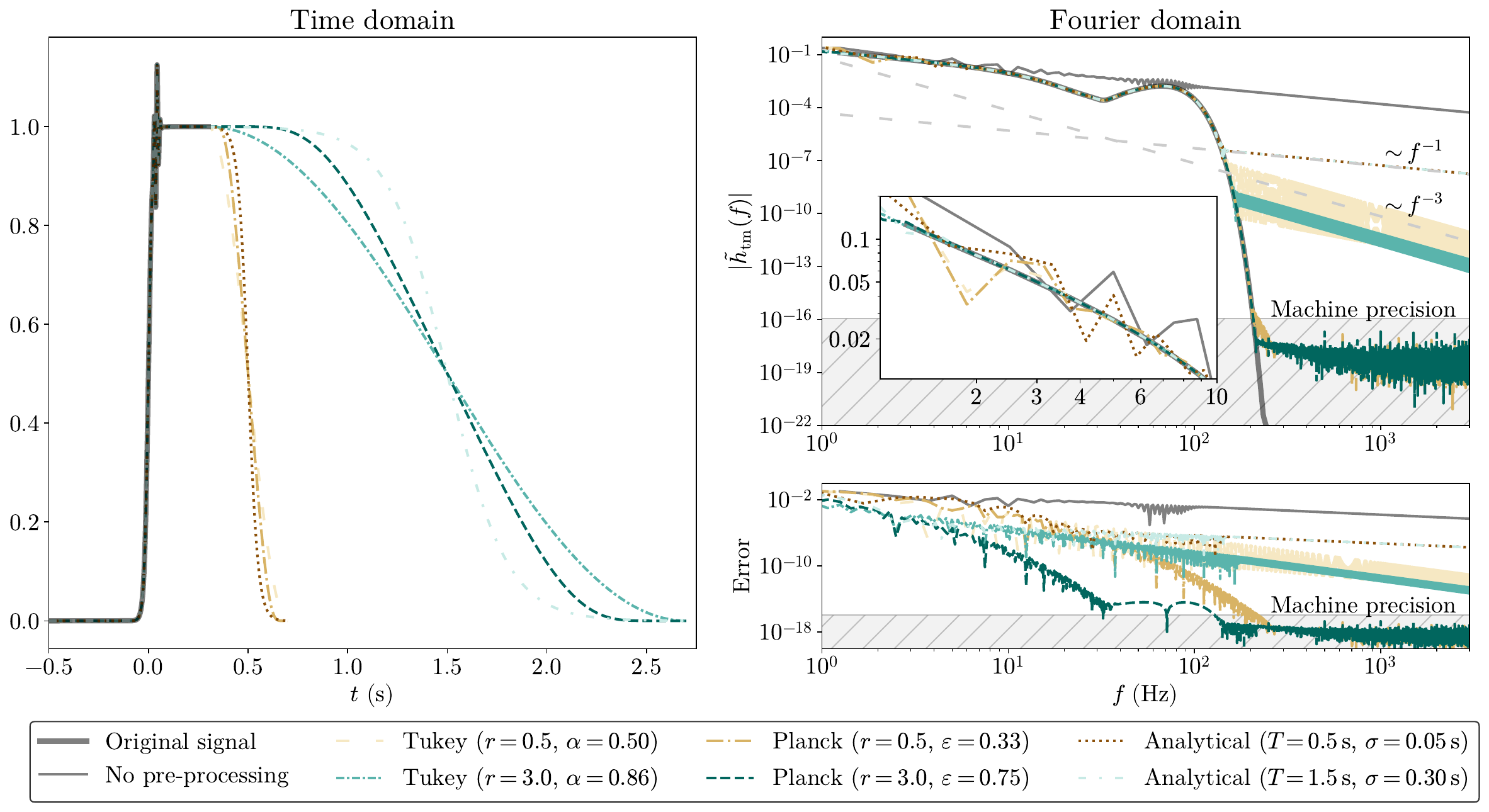}
    \caption{
        Comparison of different tapering functions applied to the GW memory toy model built with the same parameters
        as in Fig.~\ref{fig:explain_toy_model}.
        The left panel shows the resulting time-domain signal. The upper right panel shows the absolute value
        of the FT. The lower right panel shows the absolute error with respect to the closed-form result.
        The values of $r=0.5$ and $r=3.0$ for the Tukey and Planck windows correspond to
        $T=0.5\, \mathrm{s}$ and $T=1.5\, \mathrm{s}$ for the closed-form window. The number of samples of 
        the original signal (thick black line) is $N=10^6$, which is increased by a factor $(1+r)$ for the windowed signals. 
        The gray hatching approximately denotes machine precision.
    }
    \label{fig:toy_model_windows}
    \vspace{1cm}
    \includegraphics[width=\textwidth, center]{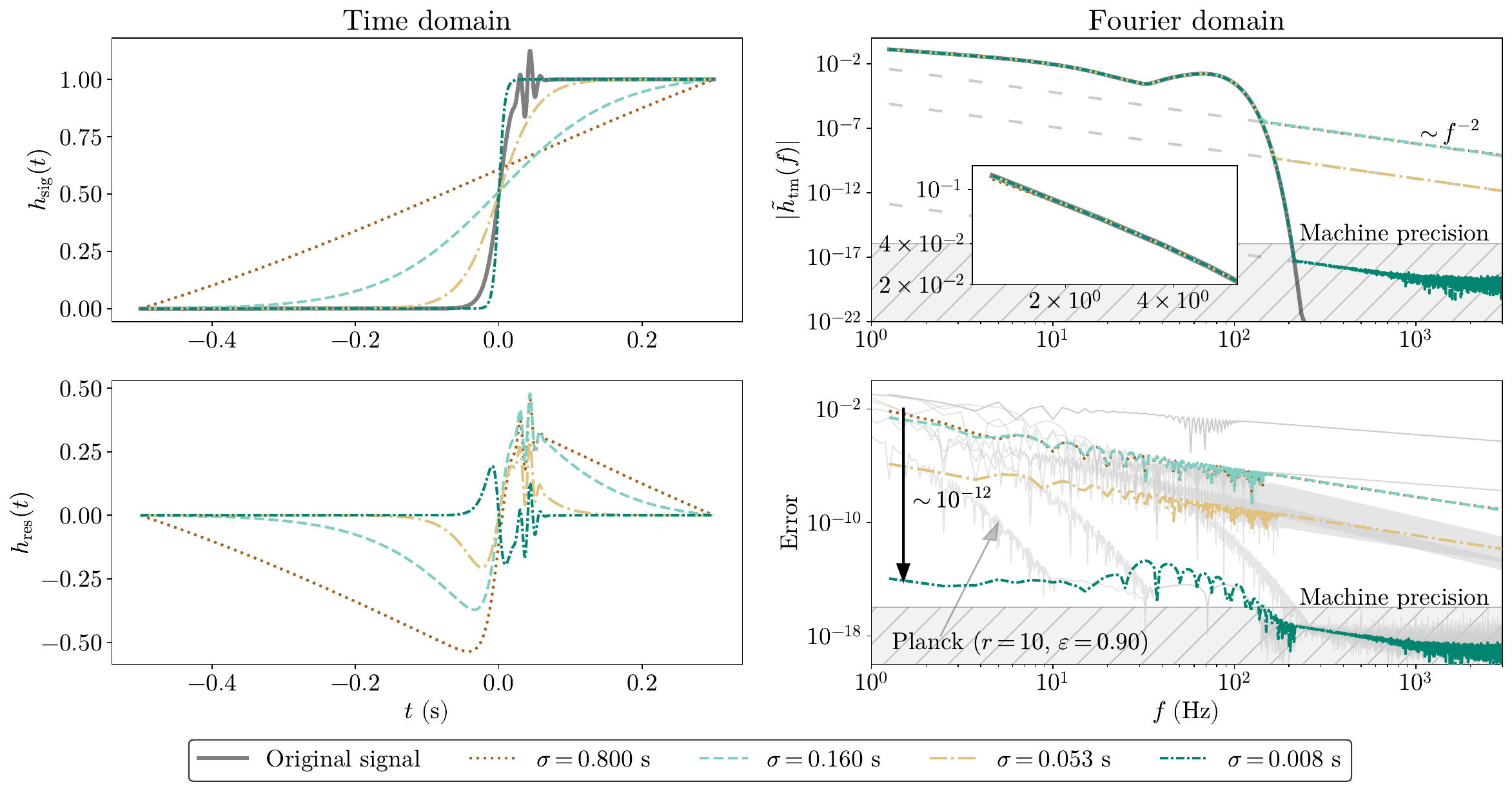}
    \caption{
        Effect of the choice of sigmoid within the \texttt{SySS} approach. The sigmoids subtracted from the toy model (upper left panel) have parameters
        $\sigma=\{0.800\,\mathrm{s},\, 0.160\,\mathrm{s},\, 0.053\,\mathrm{s},\, 0.008\,\mathrm{s} \}$ and $t_{\mathrm{jump}} = 0\,\mathrm{s}$. 
        The remaining residuals (lower left panel) share 
        the same line style as their complementary sigmoid. In the frequency spectrum (upper right panel), the dashed-gray lines follow the $\mathcal{O}(f^{-2})$
        decay of the FTs and the zoomed axis covers the region $f\in[1.15,5.9]\, \mathrm{Hz}$. The solid light-gray lines in the error plot (lower right panel) 
        are the same error curves as in the lower right panel of Fig.~\ref{fig:toy_model_windows}, except for the Planck window with $(r=10, \varepsilon=0.90)$ 
        pointed out with a gray arrow. In the low-frequency range, the black arrow emphasizes the difference of about 12 orders
        of magnitude between \texttt{SySS} with $\sigma=0.008\,\mathrm{s}$ and the standard windowing configurations.
    }
    \label{fig:toy_model_sigmoids}
\end{figure*}

\subsection{Windowing\label{subsec:windowing}}

DFTs operate under the assumption that a dataset is periodic. 
Step-like components, as discussed in Sec.~\ref{sec:practicalities_ft},
are 
interpreted as discontinuities by the DFT and give rise to 
accidental $\mathcal{O}(f^{-1})$ artifacts. A popular 
approach~\cite{Gasparotto_2023, Richardson_2022} is 
to smoothly zero the data using a window function. 
This returns a periodic dataset, thus removing the undesired artifact. 
As discussed in Sec.~\ref{sec:practicalities_ft}, however, the effect of a window 
function may interfere with that of the persistent memory, thus washing
away the signal we were interested in the first place. 

We show in Fig.~\ref{fig:toy_model_windows} an example application of windowing
to the memory toy model. We consider the ``Tukey''~\cite{1455106,tukey} and 
``Planck''~\cite{McKechan:2010kp} windows, which enjoy some popularity in the GW 
literature (see Appendix~\ref{app:windows} for expressions), plus the closed-form
window given in Eq.~\eqref{eq:td_window}. We quantify the constant padding with $r$,
which is the ratio between the number of padding samples and the original data samples.

At high frequencies, the windowed FT decays according to the window's properties:
the Tukey window is twice differentiable, so it decays as $\mathcal{O}(f^{-3})$,
while the Planck window is smooth and decays much faster.
The window defined in Eq.~\eqref{eq:td_window} does not reach 0 in a finite time;
the data is thus not exactly periodic and the decay is $\mathcal{O}(f^{-1})$. Since for a discrete setup the computational domain is finite, these
decays, based on Eq.~\eqref{eq:frequency_decays}, are only approximate and are cut off at $f_{\mathrm{samp}}/2$.

The error with respect to the true FT grows towards low frequencies. 
As discussed in Sec.~\ref{sec:practicalities_ft}, the critical frequency
up to which the window's step behavior dominates is inversely proportional
to the window's decay timescale. Windows with timescales comparable to that
of the memory, thus, return higher errors than longer windows. For the example
in Fig.~\ref{fig:toy_model_windows}, we find that a window with three times the duration
of the original signal is required to obtain an FT with an acceptable error.
Incidentally, using longer windows will also increase the computing cost
of this method.

\subsection{Symbolic Sigmoid Subtraction}\label{subsec:splitting}

In this section we present \textit{Symbolic Sigmoid Subtraction} (\texttt{SySS}), 
a simple time-domain pre-processing algorithm that  \emph{subtracts} a sigmoid function
from the data to treat the step-like behavior symbolically. This method does not 
rely on any kind of windowing and has a negligible computational overhead. 

The key idea is that all of the undesired artifacts shown in Fig.~\ref{fig:show_off}
are caused by attempting to compute the DFT of a step-like signal using a finite
data stream. As discussed in Sec.~\ref{sec:practicalities_ft}, the DFT is just an 
approximation to the continuous FT; thus, if the step-like behavior is subtracted
in closed form, its FT can be computed in the continuum and directly added,
artifact-free, to the numerical DFT of the residual signal.

In this work we choose the sigmoid function to be a hyperbolic tangent
\begin{equation}\label{eq:fitted_sigmoid}
    h_{\mathrm{sig}}(t; A, A_{\mathrm{off}}, t_{\mathrm{jump}}, \sigma) 
    = \frac{A}{2}\left[\tanh\left(\frac{t-t_{\mathrm{jump}}}{\sigma}\right) + 1 \right] + A_{\mathrm{off}} \,.
\end{equation}
The \texttt{SySS} method removes a sigmoid from the data to obtain a \emph{residual}
$h_{\mathrm{res}} = x - h_{\mathrm{sig}}$. The residual is free of step-like behaviors
(as long as $h_{\mathrm{sig}}$ has been chosen appropriately),
so $\tilde{h}_{\mathrm{res}}$ can be seamlessly computed using a DFT. 
The symbolic FT of $h_{\mathrm{sig}}$ (omitting DC components) is given
e.g. in Eq.~\eqref{eq:toy_fd_step} and only needs to be multiplied by a complex phase to match
the initial time $t_0$ of the numerical data. With this, the artifact-free FT of the data 
$\tilde{x}$ at a resolved frequency $f_k$ is given by
\begin{equation}
    \tilde{x}(f_k) = \tilde{h}_{\mathrm{res}\,k} +  e^{i \, 2 \pi f_k t_0} \tilde{h}_{\mathrm{sig}}(f_k) \,.
\end{equation}
The parameters $A$ and $A_{\mathrm{off}}$ must be chosen so that $h_{\mathrm{sig}}$
matches the persistent memory offset at the ends of the dataset, and provide the additional freedom of making the average value vanish, so that $\delta$-distributions do not appear in the Fourier transform.
The parameters  $t_{\mathrm{jump}}$ and $\sigma$, on the other hand, should be chosen such that
the ``ramp up'' of $h_{\mathrm{sig}}$ does not coincide with the time boundaries.

We show the results of applying \texttt{SySS} to the GW memory toy model in Fig.~\ref{fig:toy_model_sigmoids}.
The resulting frequency spectrum matches the expected result down to machine precision
\emph{across the whole frequency band}. In this example, this corresponds to an error reduction of up to
\emph{twelve orders of magnitude} with respect to windowing methods shown in Fig.~\ref{fig:toy_model_windows}.

The details of the behavior of \texttt{SySS} depends on the value chosen for $\sigma$:
Low $\sigma$ produces faster transitions into the asymptotic value towards the edges
of the time domain. This produces smoother residuals, which as a result tend to decay faster.
If the transition is too fast, however, step-like artifacts may be accidentally introduced
into the residual. These would cause again $\mathcal{O}(f^{-1})$ decays, albeit with a much 
lower amplitude than the original step. This can however easily be avoided. 
In addition, future models similar to \cite{Rossello-Sastre:2024zlr} could just incorporate 
the step-function decomposition in the formulation of the model.

\begin{figure}
    \centering
    \includegraphics[width=\columnwidth]{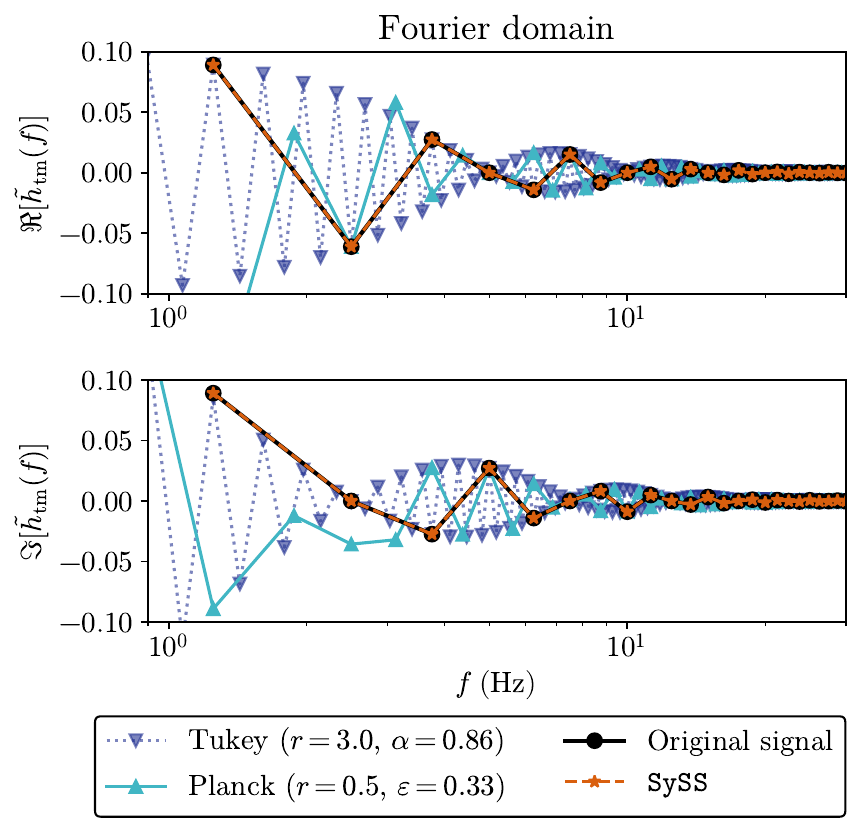}
    \caption{
        Low-frequency behavior of the real (upper panel) and imaginary (lower panel) part 
        of the GW memory toy model. The analytical values (solid line + dots) are 
        obtained from Eq.~\eqref{eq:toy_model}, while the \texttt{SySS} results (dashed-line + stars) correspond to the sigmoid with 
        $\sigma =0.008\, \mathrm{s}$ in Fig.~\ref{fig:toy_model_sigmoids}.
    }
    \label{fig:recovery_with_toy_model_re_imag}
\end{figure}

In Fig.~\ref{fig:recovery_with_toy_model_re_imag} we show the real and imaginary 
parts of the toy model's FT together with the results obtained using \texttt{SySS} 
and two windowing configurations. We note the excellent agreement of the \texttt{SySS}
result. As expected, the use of windowing spreads the signal's power into neighboring 
frequencies, which may cause issues with the detection and characterization of said 
signal~\cite{Jaynes1982OnTR,Bretthorst1988Bayesian,Allen:2002bp,Talbot:2021igi}.

\subsection{Timing comparison\label{subsec:timing}}

\begin{figure}
    \centering
    \includegraphics[width=\columnwidth]{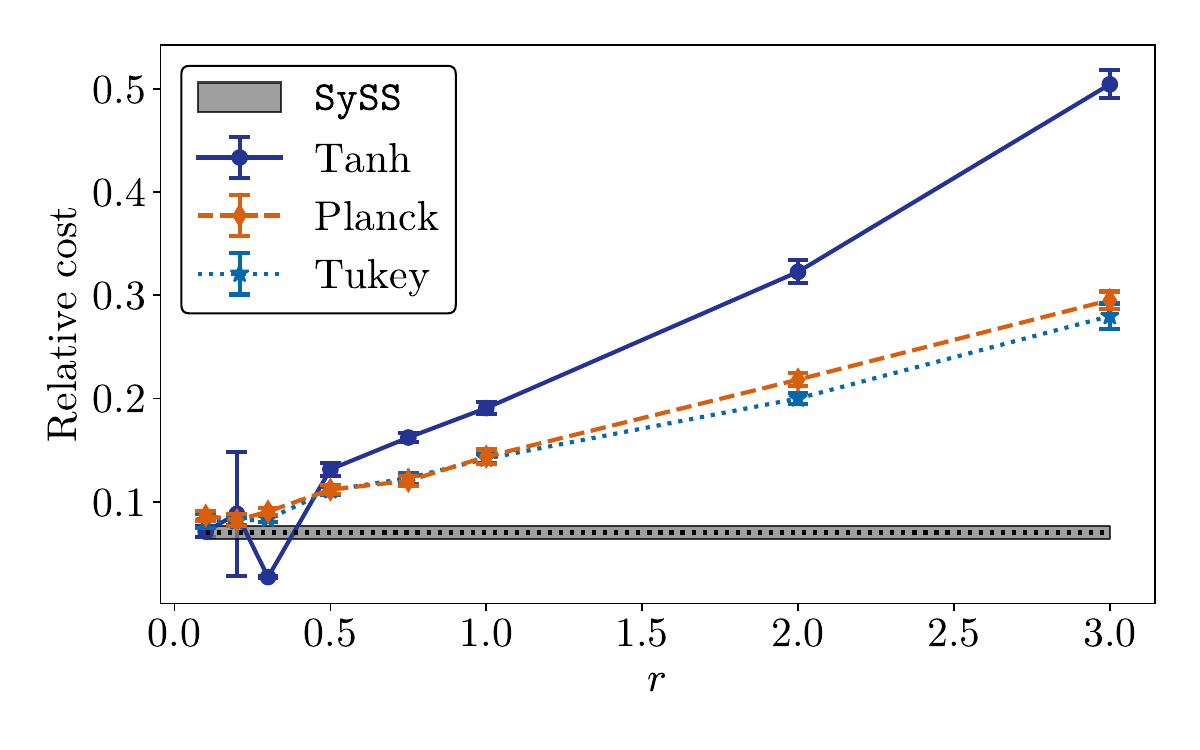}
    \caption{
        Computational cost of different pre-processing schemes relative to a fiducial waveform evaluation cost of 10 ms. 
        Tanh (solid line + dot) results are obtained by computing the DFT after windowing with Eq.~\eqref{eq:td_window}. 
        For the Planck case (dashed line + diamond), we use our own implementation based on~\cite{McKechan:2010kp} and
        for the Tukey window (dotted line + star), we use the implementation in \texttt{scipy}~\cite{2020SciPy-NMeth}.
        The \texttt{SySS} method is represented with a dotted line in a shaded area.
        We repeat each technique 1000 times and compute the average time and standard deviation. The latter
        corresponds to the error bars for the windows and the height of the shaded area for \texttt{SySS}. 
        All timings have been executed in a M3 Pro CPU.
    }
    \label{fig:relative_timings}
\end{figure}

The high number of likelihood evaluations required in GW data analysis
makes waveform generation a computationally critical step.
For time-domain waveforms, this includes the computation of the FT, as
discussed in this work. We show in Fig.~\ref{fig:relative_timings} a comparison
of the computing cost of \texttt{SySS} versus windowing strategies. 
Timings are expressed as a fraction of \SI{10}{\milli\second}, which is the average
waveform evaluation time of \texttt{IMRPhenomTHM} using \texttt{ChooseTDWaveform}
for an align-spin system with 
$\mathrm{q}=3$, $\chi_1=0.5$,  $\chi_2=-0.3$ and $\mathrm{M}=100\, \mathrm{M}_\odot$
from $f_{\mathrm{min}}=\SI{10}{\hertz}$  to $f_{\mathrm{max}}=\SI{2048}{\hertz}$ 
with a sampling rate of \SI{4096}{\hertz} \cite{Estelles:2020twz}.
This corresponds to approximately $15,000$ data points. 
Note that this is
an illustrative example, since results will depend on the application and waveform model used.

The cost of windowing depends on the window's length, which here we parametrize
as the ratio $r$ between the window's and original data's length. 
Longer windows incur a higher computing cost as the FT to be computed involves a higher number of data points.
\texttt{SySS}, on the other hand, only evaluates closed-form expressions and does not extend
the duration of the dataset. Its cost is thus constant and about 7\% of the evaluation 
time of the waveform. We remark that \texttt{SySS} only involves the subtraction and addition of closed-form
expressions to an array with a length given by the specific application. It is thus expected that the computing
cost of this method will not scale strongly with longer signals, as the only computationally-expensive operation
is the computation of a closed-form expression.

As shown previously in Fig.~\ref{fig:toy_model_sigmoids}, window functions must approach zero
rather slowly to avoid interfering with the memory signal. In our examples, this corresponds
typically to $r \gtrsim 3$. For the Tukey and Planck windows, the corresponding computing cost, shown in
Fig.~\ref{fig:relative_timings}, is $\sim 30\%$ of the waveform's evaluation time.

This comparison establishes \texttt{SySS} as a better approach, in the sense that it is able
to return the exact FT at a smaller computing cost with no tuning required. This is because, rather
than attempting to suppress a numerical artifact, \texttt{SySS} directly treats the root cause of the problem,
namely a step-like behavior in a discrete time series, by properly treating the problem in the continuum.   
\section{Phenomenology of Fourier domain GW memory signals\label{sec:memory}}

We now turn to the treatment of actual gravitational-wave signals. The gravitational-wave strain $h$ as emitted by a binary system has 
two independent polarisations, $h_+$ and $h_\times$, which depend on 
the inertial time coordinate $t$, the distance to source $d_\mathrm{L}$, the intrinsic 
physical parameters of the source $\bm{\lambda}$, and source's orientation
angles $(\theta,\phi)$. These polarisations can be decomposed in a basis
of spin-weighted spherical harmonics \cite{10.1063/1.1705135, Wiaux_2007}:
\begin{align}
    h(t;d_\mathrm{L},\bm{\lambda},\theta,\phi) 
    & =  h_+(t;d_\mathrm{L},\bm{\lambda},\theta,\phi) - i h_\times(t;d_\mathrm{L},\bm{\lambda},\theta,\phi) \label{eq:td_polarisarion_decomposition} \\ 
    & = \frac{1}{d_\mathrm{L}}\sum\limits_{\ell \geq 2}\sum\limits_{|m|\leq \ell} {}^{-2}Y_{\ell m}(\theta, \phi) \, h_{\ell m}(t;\bm{\lambda})
    \label{eq:td_mode_decomposition} 
\end{align}
 where $h_{\ell m}(t;\bm{\lambda})$ are the spherical harmonic modes 
 and depend exclusively on the time and the intrinsic physical parameters
 of the source. See e.g.~\cite{Talbot:2018sgr} for a discussion of the spherical
 harmonic structure for memory waveforms.

For concreteness we focus on the  $h_{20}$ spherical harmonic, which contains the dominant 
gravitational-wave memory contribution, however, other spherical harmonics could be treated with the same methods.
The $h_{20}$ 
signal has two components, namely a non-oscillatory growth caused 
by the gravitational-wave memory effect $h_{20}^{(\text{mem})}$,
and an oscillatory contribution from the quasi-normal 
ringdown $h_{20}^{(\text{osc})}$: 
\begin{equation}\label{eq:20_mem_ans_osc}
    h_{20}(t;\bm{\lambda}) = h_{20}^{(\text{mem})}(t;\bm{\lambda}) + h_{20}^{(\text{osc})}(t;\bm{\lambda}).
\end{equation}
The memory contribution persists after the gravitational-wave transient has passed and leaves 
a constant offset in the signal. This makes the amplitude to be non-zero after the merger-ringdown, 
which triggers artifacts in the Fourier-transformed waveform, as discussed in Sec.~\ref{sec:practicalities_ft}. 
Due to this step-like morphology, \texttt{SySS} is a suitable pre-processing technique before computing the DFT, 
as we have already seen in Sec.~\ref{sec:toy_model}. 

Throughout this section we fix $\theta = \pi/2$, for which the $h_{20}$-mode is maximal, and $\phi=0$.  For convenience, we define
\begin{equation}
    \mathfrak{h}_{\ell m } = (d_\mathrm{L}/\mathrm{M})\, ^{-2}Y_{\ell m}(\theta=\pi/2, \phi=0)\, h_{\ell m}
\end{equation}
and 
\begin{equation}
    \tilde{\mathfrak{h}}_{\ell m } =
    (d_\mathrm{L}/\mathrm{M}^2)\, ^{-2}Y_{\ell m}(\theta=\pi/2, \phi=0)\, \tilde{h}_{\ell m}
\end{equation}
so that the time-domain (frequency-domain) strain
is simply $\sum_{\ell m}\mathfrak{h}_{\ell m}$ $\left(\sum_{\ell m}\tilde{\mathfrak{h}}_{\ell m}\right)$
in geometric units. 

Whenever applying \texttt{SySS} in this section, we choose $t_{\mathrm{jump}}$ to coincide
with merger time and $\sigma = 10\, \mathrm{M}$.
As discussed in Sec.~\ref{subsec:splitting},
this choice is not unique, as \texttt{SySS} focuses on capturing the general step-like behavior
of the signal. As shown in Appendix~\ref{app:syss_choices}, results are broadly unaffected by the selection of $t_{\mathrm{jump}}$ and $\sigma$.
It is important to note that \texttt{SySS} is \emph{not} explicitly modeling $h_{20}^{\text{(mem)}}$,
but rather \emph{any} step-like behavior would be successfully dealt with.

\subsection{Global picture of $h_{20}$ in the frequency domain}\label{subsec:only_20}

\begin{figure}[t]
    \centering
    \includegraphics[width=\columnwidth]{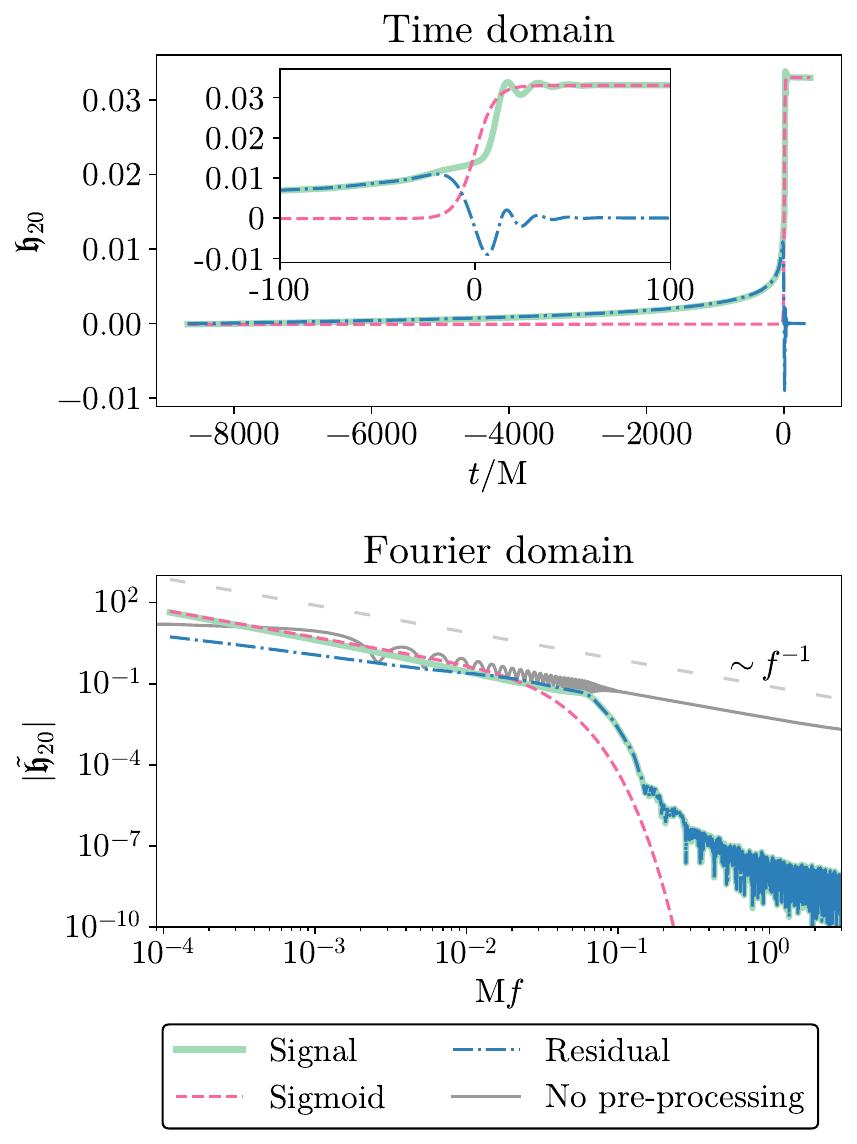}
    \caption{
        Time (upper panel) and Fourier (lower panel) domain of the (2,0) mode (thick solid line)
        for a non-spinning equal-mass binary black hole system (SXS:BBH:0001).
        The time range shown spans $t/\mathrm{M}\in[-8674.52,\, 382.20]$. 
        The dashed line represents the sigmoid used for \texttt{SySS}, while the dash-dotted line displays the residual.
        A closer look at the merger is shown in the zoomed axis for times between $\pm 100\,\mathrm{M}$.
        In the lower panel, the thin solid line represents the FT of the (2,0) mode without applying any
        pre-processing technique, and the loosely-dashed line shows the $\mathcal{O}(f^{-1})$ step-like decay.
        The memory contribution $h_{20}^{(\text{mem})}$
        shown here has been computed using the \texttt{sxs.waveforms.memory.add\_memory}
        option \cite{PhysRevD.103.024031}.
    }
    \label{fig:realistic_memory_waveform1}
\end{figure}

We start by computing the FT of a pure (2,0) NR waveform using \texttt{SySS} 
to understand the frequency-domain phenomenology of GW memory. For this, we choose an NR
waveform from the SXS catalog~\cite{SXS:catalog}, shown in Fig.~\ref{fig:realistic_memory_waveform1}.
As expected, the sigmoid component, which captures the step-like behavior, 
is dominant at low frequencies. The residual, which essentially carries the ringdown component, 
dominates at high frequencies.

\subsection{Memory and oscillatory contributions}\label{subsec:20_mem_and_osc}

In this section we apply \texttt{SySS} separately to $h_{20}^{\text{(mem)}}$ 
and $h_{20}^{\text{(osc)}}$, and we compare their frequency spectrum with the 
one from the most relevant higher harmonics, in different regions of the parameter space. 
We have computed $h_{20}^{(\mathrm{mem})}$ with Eq.~(3.27) of \cite{Rossello-Sastre:2024zlr},
while $h_{20}^{\text{(osc)}}$ and the rest of the modes have been obtained directly
from NR simulations of the SXS Catalog \cite{SXS:catalog}.

\begin{figure*}[tbph]
    \centering
    \includegraphics[width=\textwidth]{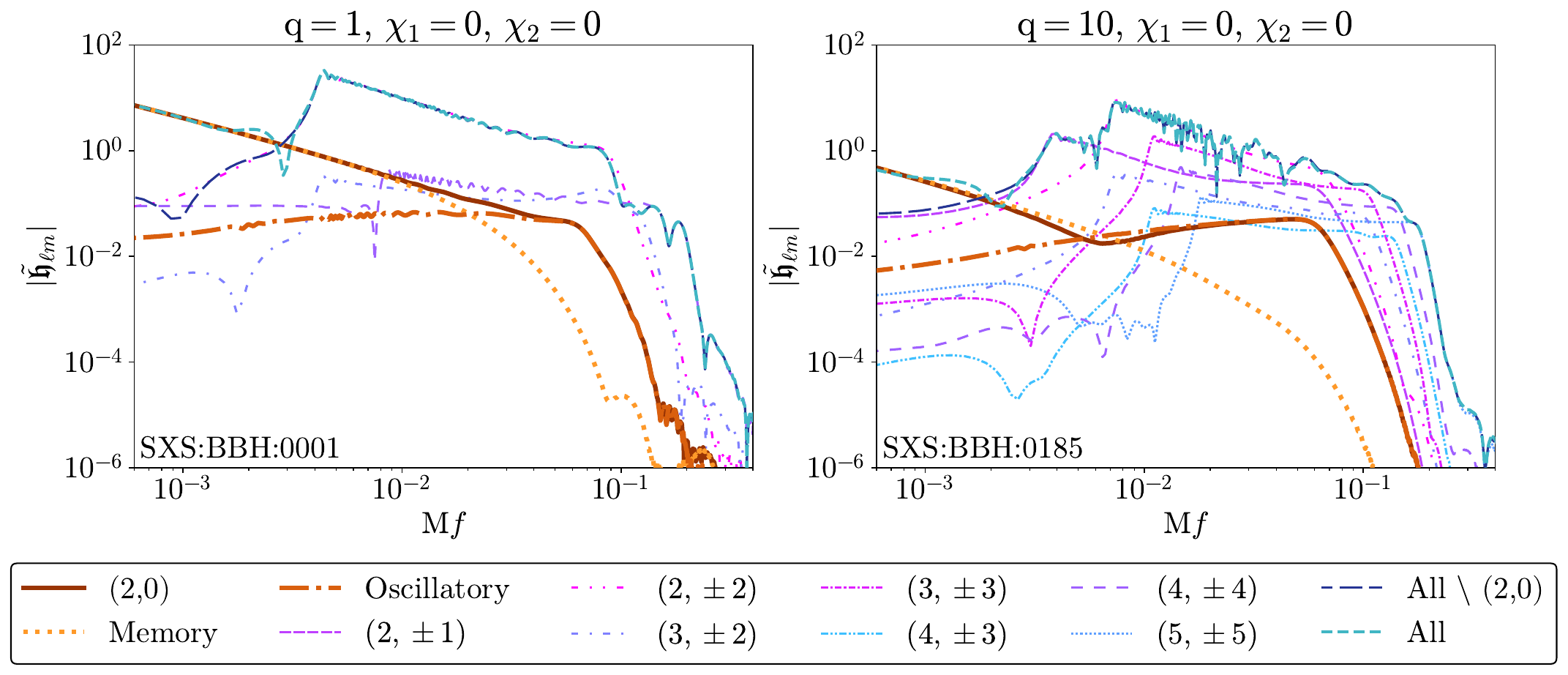}
    \caption{
        Frequency-domain decomposition of the GW into some of its multipolar contributions
        for the non-spinning systems SXS:BBH:0001 (left), and SXS:BBH:0185 (right), with mass ratios 1:1 and 1:10, respectively.
        The (2,0) mode (thick solid line) is split into the memory (thick dotted line) and the oscillatory 
        (thick dash-dotted line) contributions as in Eq.~\eqref{eq:20_mem_ans_osc}. 
        The memory contribution is computed from Eq.~(3.27) of~\cite{Rossello-Sastre:2024zlr}, while the oscillatory part
        and the rest of the modes are obtained from the corresponding NR simulations of the SXS Catalog \cite{SXS:catalog}.
        The \texttt{SySS} method, with the standard parameters $\sigma = 10\,\mathrm{M}$ and $t_{\mathrm{jump}}=t_{\mathrm{merger}}$, 
        has been only applied to $h_{20}$, $h^{(\mathrm{mem})}_{20}$ and $h^{(\mathrm{osc})}_{20}$.
    }
    \label{fig:NR_map_non_spinning}
\end{figure*}

\begin{figure*}[tbph]
    \centering
    \includegraphics[width=\textwidth]{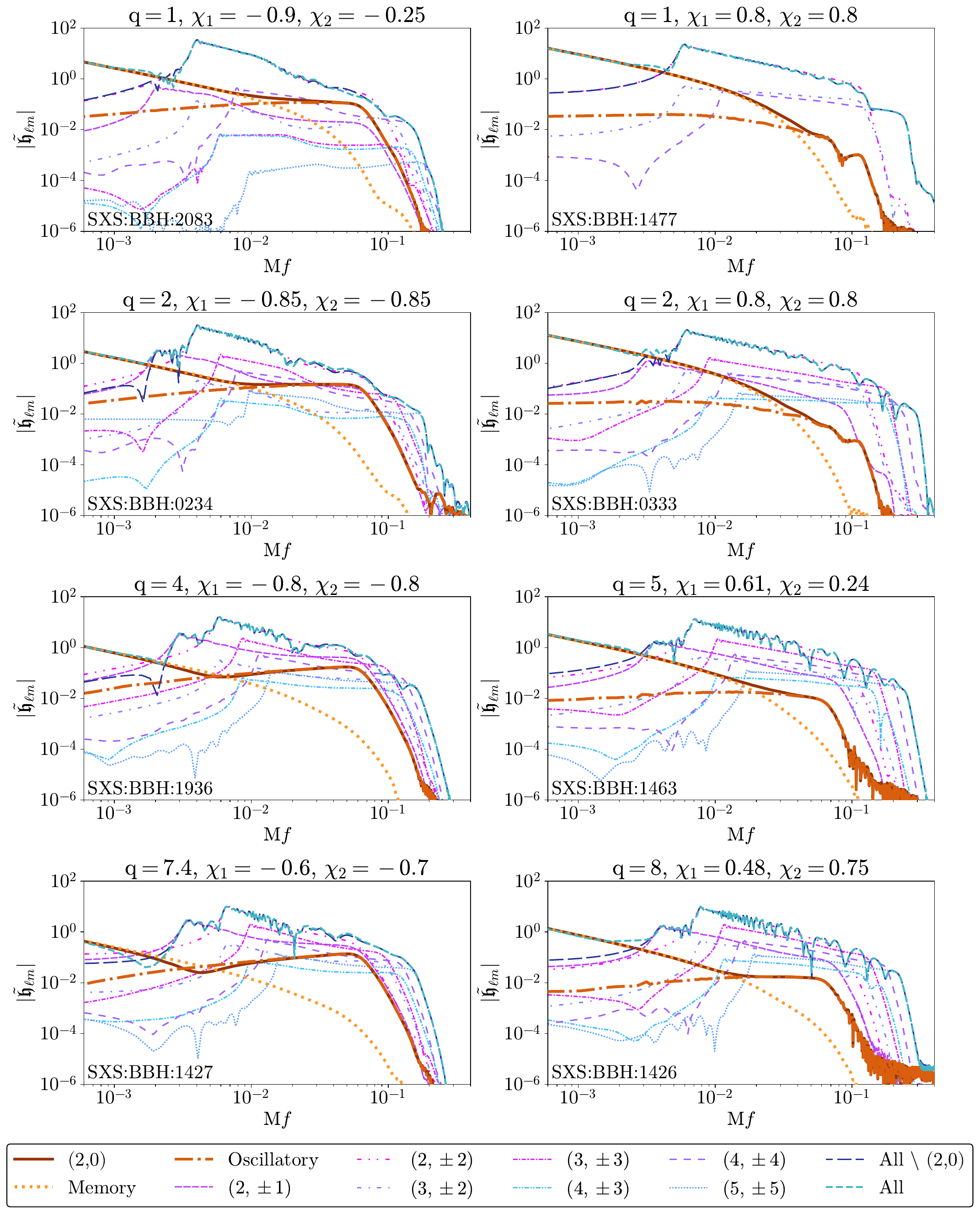}
    \caption{
        Same multipolar GW decomposition as in Fig.~\ref{fig:NR_map_non_spinning}, 
        but for the align-spin systems: 
        SXS:BBH:2083, SXS:BBH:1477, SXS:BBH:0234, SXS:BBH:0333, SXS:BBH:1936, SXS:BBH:1463, SXS:BBH:1427, SXS:BBH:1426.
        The plots are arranged such that the left column corresponds to negative-spin systems, 
        right column to positive-spin systems, and the mass ratio increases as one goes down a column.
        Furthermore, the $y$ and $x$-axes are limited to the same maximum and minimum values to facilitate comparisons between plots.
    }
    \label{fig:NR_map_spinning}
\end{figure*}

The results are shown in Figs.~\ref{fig:NR_map_non_spinning} and~\ref{fig:NR_map_spinning}. 
Note that the finite duration of NR waveforms introduces artifacts in the FT at low frequencies. This is shown, for instance, in the left panel of Fig.~\ref{fig:NR_map_non_spinning}. In the Fourier domain, the amplitude of the (2,2) mode increases toward low frequencies following a leading-order $f^{-7/6}$ power law. Once the maximum is reached, around $\rm{M} f \approx 0.004$ for this case, there is a sharp drop caused by the finite length of the time-domain signal. As a result, there is a low-frequency region where the amplitude of the (2,0) mode unphysically exceeds that of the (2,2) mode. This unrealistic behavior can be safely ignored since it does not affect the results of this work.

Overall, we note differentiated behaviors between $\tilde{h}_{20}^{(\mathrm{mem})}$
and $\tilde{h}_{20}^{(\mathrm{osc})}$ for low and high frequencies, as well
as a modification of the complete waveform when adding the (2,0) mode.

The memory contribution to the (2,0) mode is always dominant at low frequencies,
where it behaves like a step-function with the standard $\mathcal{O}(f^{-1})$ trend.
At high frequencies, however, the effect of $\tilde{h}_{20}^{(\mathrm{mem})}$ is
smaller than the contribution from the oscillatory part since it has already decayed by the time
the merger-ringdown takes place. 
This behavior is consistent with the one previously discussed in Sec.~\ref{sec:toy_model} 
with the toy model. Furthermore, we observe that the relative contribution of 
$\tilde{h}_{20}^{(\mathrm{mem})}$ to the frequency spectrum becomes larger 
for symmetric mass ratios and equal positive spins,
which agrees with the results obtained in~\cite{Rossello-Sastre:2024zlr, Xu:2024ybt, Pollney_2011}.

The oscillatory contribution is negligible during the inspiral, 
and it becomes prominent as it approaches the merger.
For the non-spinning case (Fig.~\ref{fig:NR_map_non_spinning}), 
the power coming from this contribution in the spectrum increases with the mass ratio.
This behavior is also seen for spinning systems (Fig.~\ref{fig:NR_map_spinning}). 
In addition, we find that $\tilde{h}_{20}^{(\mathrm{osc})}$ has a greater impact for 
high-negative spins, whereas it is reduced for positive spins and even more for equal positive spins, 
which is consistent with \cite{Rossello-Sastre:2024zlr, Pollney_2011}. 

Including the (2,0) mode spherical harmonic in Eq.~\eqref{eq:td_mode_decomposition} adds a $\mathcal{O}(f^{-1})$ contribution at low frequencies.
For larger frequencies, it can significantly modify the merger-ringdown part of the full waveform, 
especially for high negative spins, where the oscillatory part is louder than most of the higher harmonics
(see left panels of Fig.~\ref{fig:NR_map_spinning}).

\subsection{\texttt{SySS} all at once}\label{subsec:all_modes}

The \texttt{SySS} method can be applied to compute the FT of the full waveform at once.
This is because we are not trying to subtract \emph{the} step, but rather \emph{a} step.
We show an example case in Fig.~\ref{fig:sys_all_at_once}, which demonstrates the 
robustness of the method. We observe a perfect agreement between two independent methods. 
The first one consists of computing the FT of the waveform mode by mode, 
where \texttt{SySS} is only applied to the (2,0) mode, 
whereas the remaining FTs are computed tapering at the early and late times. 
The second method, however, applies the \texttt{SySS} technique to the whole waveform, since it has 
a permanent offset originated by the contribution of the (2,0) mode.

\section{Discussion}\label{sec:discussion}

GW data analysis for current ground-based detectors is typically conducted in the frequency domain.
This requires a careful treatment of GW memory, which consists of step-like time-domain signals and has motivated the development of multiple data-analysis recipes to suppress undesired numerical artifacts. 
These recipes, as we show in Sec.~\ref{sec:toy_model}, critically distort low-frequency components, 
where the memory contribution is more prominent, and thus may hinder a first GW memory detection 
in future ground-based detectors, for which detection prospects are more favorable.
 
In this work, we presented \texttt{SySS}, an embarrassingly simple algorithm to treat generic step-like signals in the time domain using closed-form FTs. This new method treats step-like contributions
in the continuum to avoid numerical artifacts.

We have shown the imprint the (2,0) mode leaves on the full frequency-domain waveform.
At low frequencies, it is characterized by a $\mathcal{O}(f^{-1})$ trend coming from the step component,
while for high frequencies the oscillatory contribution can introduce deviations to
the merger-ringdown, especially for high-negative spin systems.
We found a more prominent contribution from the oscillatory part as mass ratio 
increases and for high-negative spin systems,  whereas the memory contribution
dominates for low mass ratio and positive spins. Those results are consistent 
with~\cite{Pollney_2011} and the recent work done in~\cite{Rossello-Sastre:2024zlr, Xu:2024ybt}. 

Accuracy and timing comparisons versus windowing approaches conclude that \texttt{SySS}
is able to match the expected FT of a step-like signal down to numerical precision at a 
negligible increase in computing cost. This makes \texttt{SySS} a particularly useful method to construct Fourier domain GW memory waveforms.

We release an open-source Python implementation of \texttt{SySS}~\cite{FouTStep}
which can be readily hooked up with the LALSuite waveform interface.

This work will help in developing future Fourier-domain models for the (2,0) mode, which can be constructed in a way that is independent of any windowing techniques.
This helps in particular to develop a clear understanding of the morphology of
memory signals in the frequency domain.

\section*{Acknowledgements}

Jorge Valencia is supported by the Spanish Ministry of Universities (FPU22/02211),
Maria Rossell\'o-Sastre is supported by the Spanish Ministry of Universities (FPU21/05009).
Rodrigo Tenorio is supported by 
ERC Starting Grant No.~945155--GWmining, 
Cariplo Foundation Grant No.~2021-0555, 
MUR PRIN Grant No.~2022-Z9X4XS, 
MUR Grant ``Progetto Dipartimenti di Eccellenza 2023-2027'' (BiCoQ),
and the ICSC National Research Centre funded by NextGenerationEU.
This work was supported by the Universitat de les Illes Balears (UIB); the Spanish Agencia Estatal de Investigaci\'on grants PID2022-138626NB-I00, RED2022-134204-E, RED2022-134411-T, funded by MICIU/AEI/10.13039/501100011033, by the ESF+ and ERDF/EU; the MICIU with funding from the European Union NextGenerationEU/PRTR (PRTR-C17.I1); the Comunitat Aut\`onoma de les Illes Balears through the Direcci\'o General de Recerca, Innovaci\'o I Transformaci\'o Digital with funds from the Tourist Stay Tax Law (PDR2020/11 - ITS2017-006), the Conselleria d’Economia, Hisenda i Innovaci\'o grant numbers SINCO2022/18146 and SINCO2022/6719, co-financed by the European Union and FEDER Operational Program 2021-2027 of the Balearic Islands.
This paper has been assigned document number LIGO-P2400263.

\appendix

\section{Shifting in time domain}\label{app:shift}

For some applications, it may be convenient to circularly time-shift the finite dataset
to reduce the oscillations in the frequency domain, according to Eq.~\eqref{eq:FT_shift_property}.

For the Gaussian example of Sec.~\ref{subsec:gaussian_example}, the oscillatory behavior introduced by the phase factor of 
Eq.~\eqref{eq:ft_gaussian} can be countered by choosing a time shift $t_\mathrm{s}=-\vert t_0 - \mu \vert$, which also takes into account the oscillatory component
introduced by a non-zero $t_0$, as discussed in Sec.~\ref{subsec:dft}. In general, for localized signals, $t_\mathrm{s}$ must be
such that the dynamic range of the data is split half at the beginning of the domain and half at the end.

\begin{figure*}[!t]
    \centering
    \includegraphics[width=\textwidth]{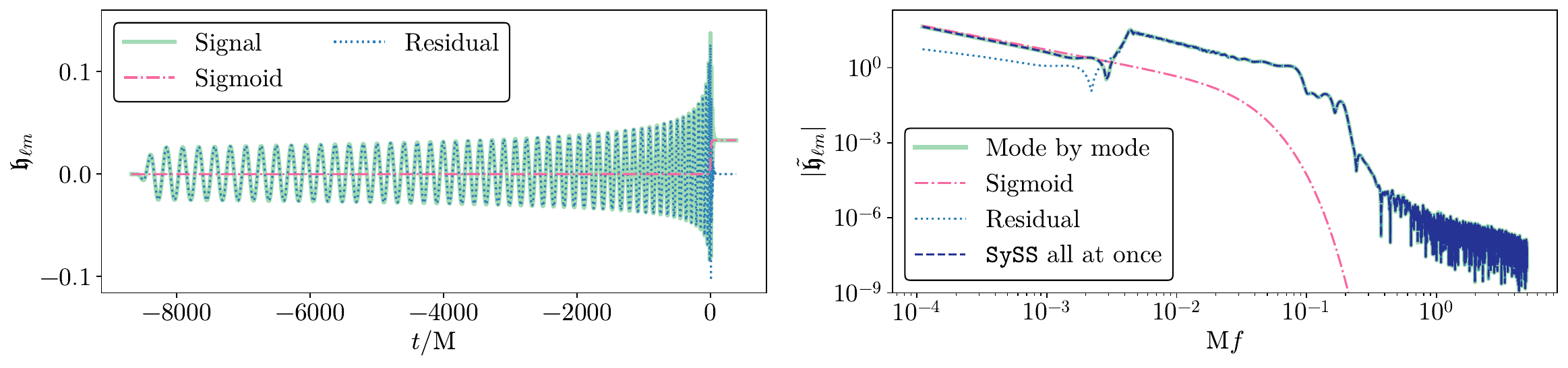}
    \caption{
        \texttt{SySS} technique applied to the GW from the equal-mass non-spinning system SXS:BBH:0001, where we
        have taken into account the same spherical harmonic modes shown in Figs.~\ref{fig:NR_map_non_spinning} and~\ref{fig:NR_map_spinning}.
        The left panel displays the time-domain signal (thick solid line) as well as the sigmoid (dash-dotted line) and
        residual (dotted line) used in \texttt{SySS}. The parameters of the sigmoid are the same as in Fig.~\ref{fig:NR_map_non_spinning}.
        The right panel compares the FT obtained by summing the FT of the modes one by one (thick solid line), as done in Figs.~\ref{fig:NR_map_non_spinning} 
        and~\ref{fig:NR_map_spinning}, with the FT obtained by applying \texttt{SySS} to all the modes at once (dashed line).
    }
    \label{fig:sys_all_at_once}
\end{figure*}

On the other hand, for not well-localized signals, a suitable option can be $t_\mathrm{s}=-\vert t_0 - t_{\mathrm{max}} \vert$, 
being $t_{\mathrm{max}}$ the time where the maximum of the signal takes place.

\begin{figure*}[ptb]
    \centering
    \includegraphics[width=\textwidth]{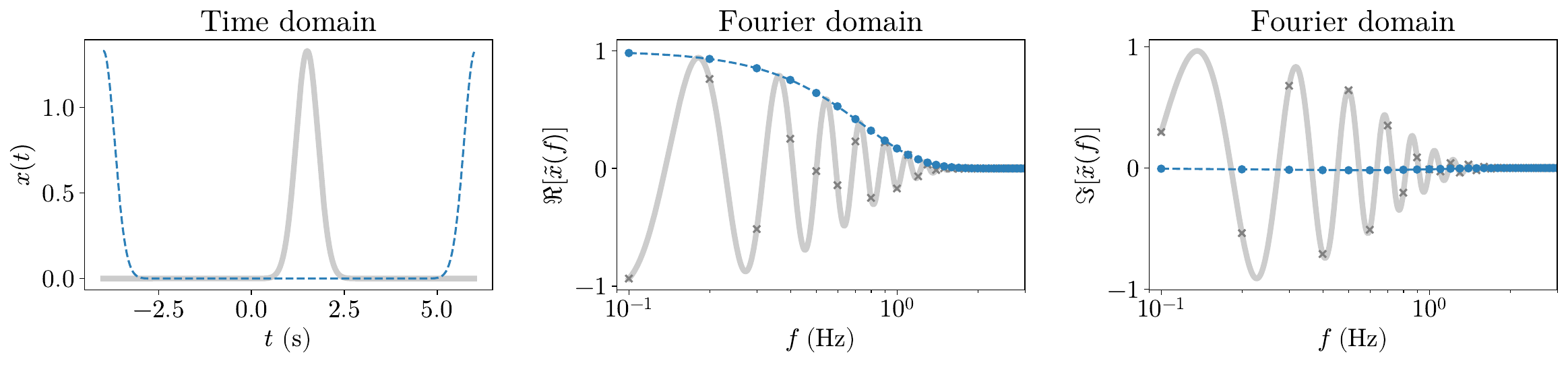}
    \caption{
        Effect of a time shift in the FT of a Gaussian with parameters $\mu=1.5\,\mathrm{s}$ and $\sigma=0.3\, \mathrm{s}$.
        The left panel shows the time-domain unshifted Gaussian profile (thick solid line) and the same profile after a time shift of
        $t_{\mathrm{s}}=-\vert t_0 - \mu \vert$ (dashed line). The time grid starts at $t_0=-4\,\mathrm{s}$ and samples $N=500$ points with a constant time step $\delta t = 0.02\, \mathrm{s}$. 
        The middle and right panels represent the real and imaginary parts of the continuous (lines) and discrete (markers) FTs. The 
        discrete version has been evaluated at multiples of the Rayleigh frequency $1/(N\delta t)$.
    }
    \label{fig:shifting_around}
\end{figure*}

\begin{figure*}[ptb]
    \centering
    \includegraphics[width=\textwidth]{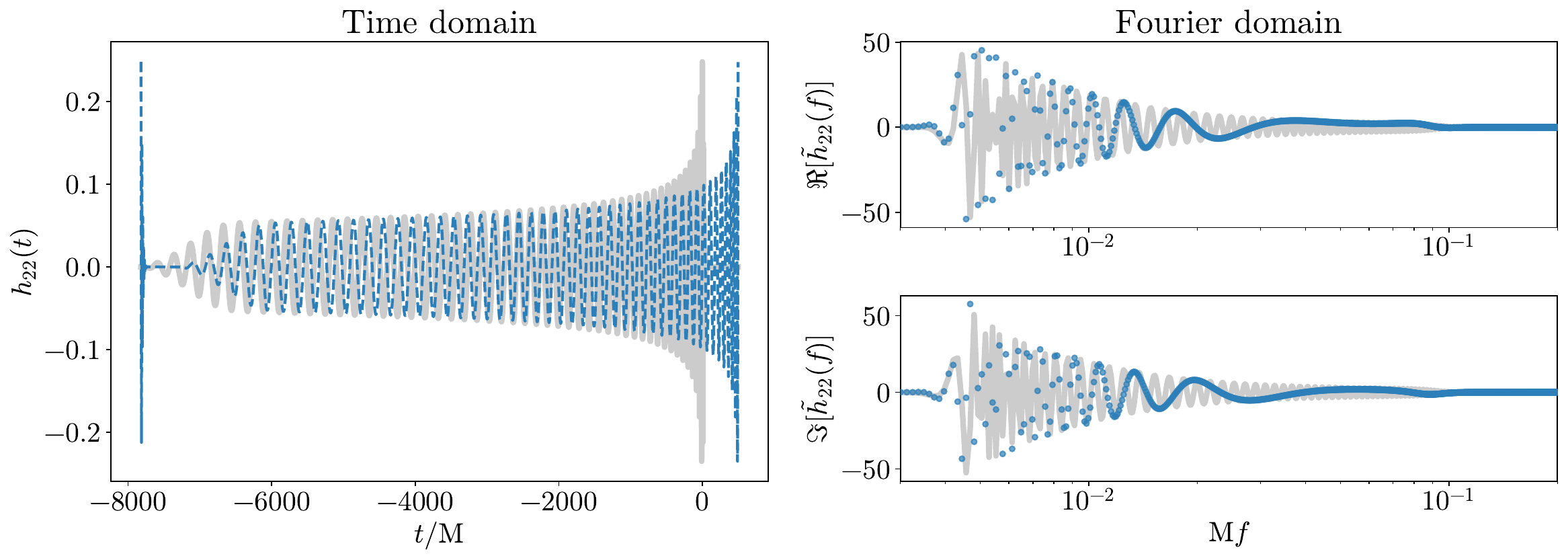}
    \caption{
            Same as in Fig.~\ref{fig:shifting_around} but for a GW signal for a non-spinning equal-mass system.
            The waveform has been generated with the LALSuite software infrastructure using the approximant \texttt{IMRPhenomTHM} \cite{Estelles:2020twz}
            including only the (2,2) mode. Without loss of generality, we have aligned the waveform at the peak so that $t_{\mathrm{max}}=0$.
            Unlike in Fig.~\ref{fig:shifting_around}, no continuous FT is displayed here.
    }
    \label{fig:shifting_around_22waveform}
\end{figure*}

We compare the effect of these time shifts for a Gaussian profile (Fig.~\ref{fig:shifting_around}) and a GW signal 
(Fig.~\ref{fig:shifting_around_22waveform}). In the first case, the well location of the signal allows to remove the oscillatory 
components of the FT efficiently, whereas for the GW example, since there is not a well-established location, only a partial
counteracting of the oscillations is achieved, mostly at high frequencies.

\section{Window functions}\label{app:windows}

We define discrete window functions as a time series with $N$ samples labeled by
$j = 0, \dots, N-1$ regardless of its initial time $t_0$. Both of the windows
here considered are parameterised by a single parameter which tunes the transition
length between 0 and 1.

The Tukey window is defined as
\begin{align}
    w_j &= 
    \frac{1}{2}\left[ 1 - \cos\left(\frac{\pi}{\Delta(\alpha)}j\right)\right]
    &0\leq j < \Delta(\alpha)\\
    w_j &= 1
    &\Delta(\alpha) \leq j \leq (N-1)/2 \\
    w_{j} &  = w_{(N-1) - j} & (N-1)/2 < j \leq N-1
\end{align}
where $\Delta(\alpha) = \alpha (N-1)/2$ is the transition interval and we assume
integer division. This window is only twice differentiable.

The Planck window is defined as
\begin{align}
    w_j & = 0 & j = 0\\
    w_j & = \left[ 1 + e^{\left( \frac{\Delta(\varepsilon) }{j} 
    - \frac{\Delta(\varepsilon)  }{\Delta(\varepsilon)  - j} \right)} \right]^{-1} & 1 \leq j < \Delta(\varepsilon) \\
    w_j & = 1 & \Delta(\varepsilon) \leq j \leq (N - 1)/2 \\
    w_{j} &  = w_{(N-1) - j} & (N-1)/2 < j \leq N-1
\end{align}
where $\Delta(\varepsilon) = \varepsilon(N-1)$. This window is based on a 
bump function, which is smooth. 
\section{The choice of $t_{\rm{jump}}$ and $\sigma$ in \texttt{SySS}}\label{app:syss_choices}

\begin{figure*}[ptb]
     \centering
     \begin{subfigure}[b]{0.495\textwidth}
         \centering
         \includegraphics[width=\textwidth]{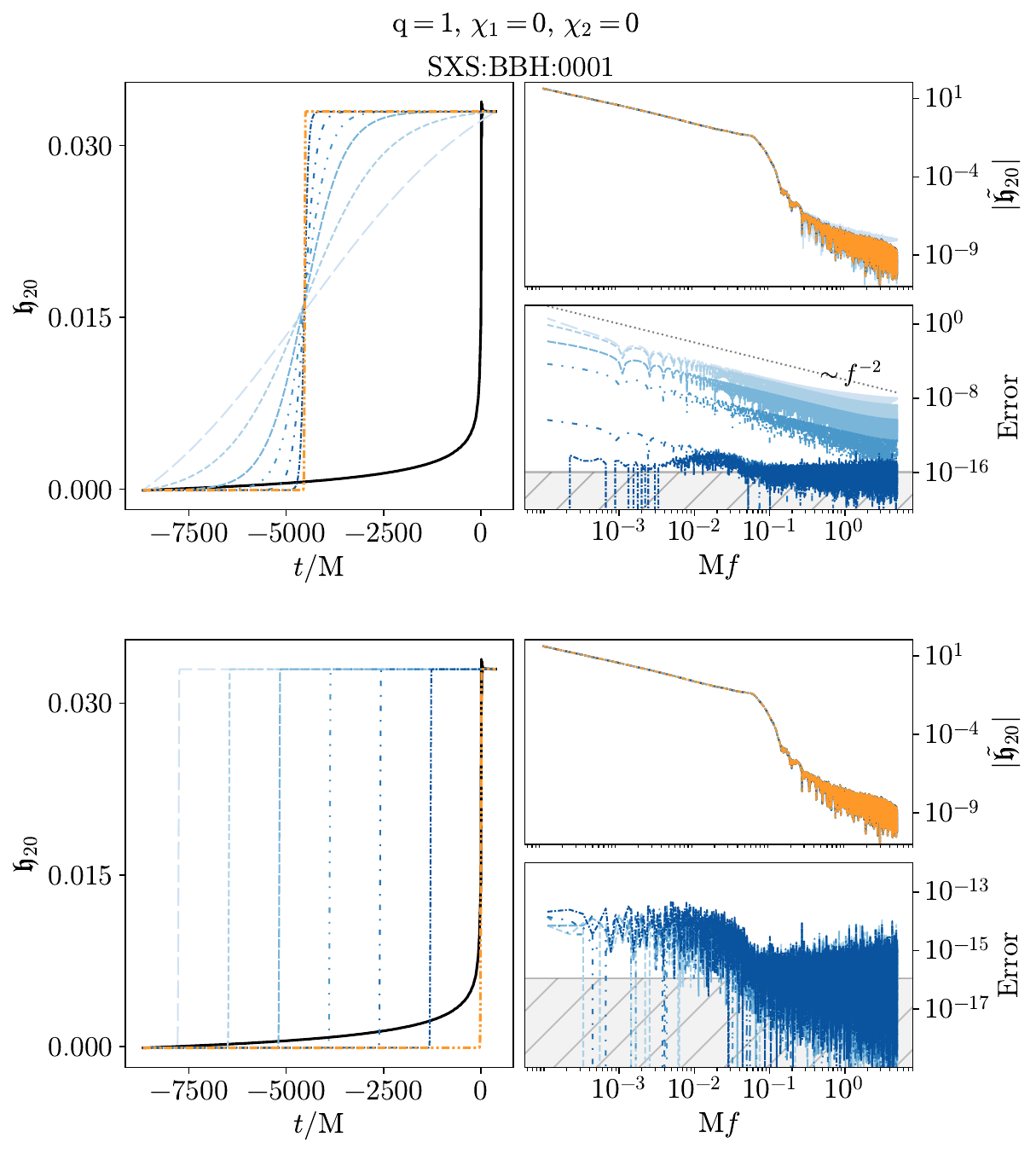}
         \label{fig:y equals x}
     \end{subfigure}
     \hfill
     \begin{subfigure}[b]{0.495\textwidth}
         \centering
         \includegraphics[width=\textwidth]{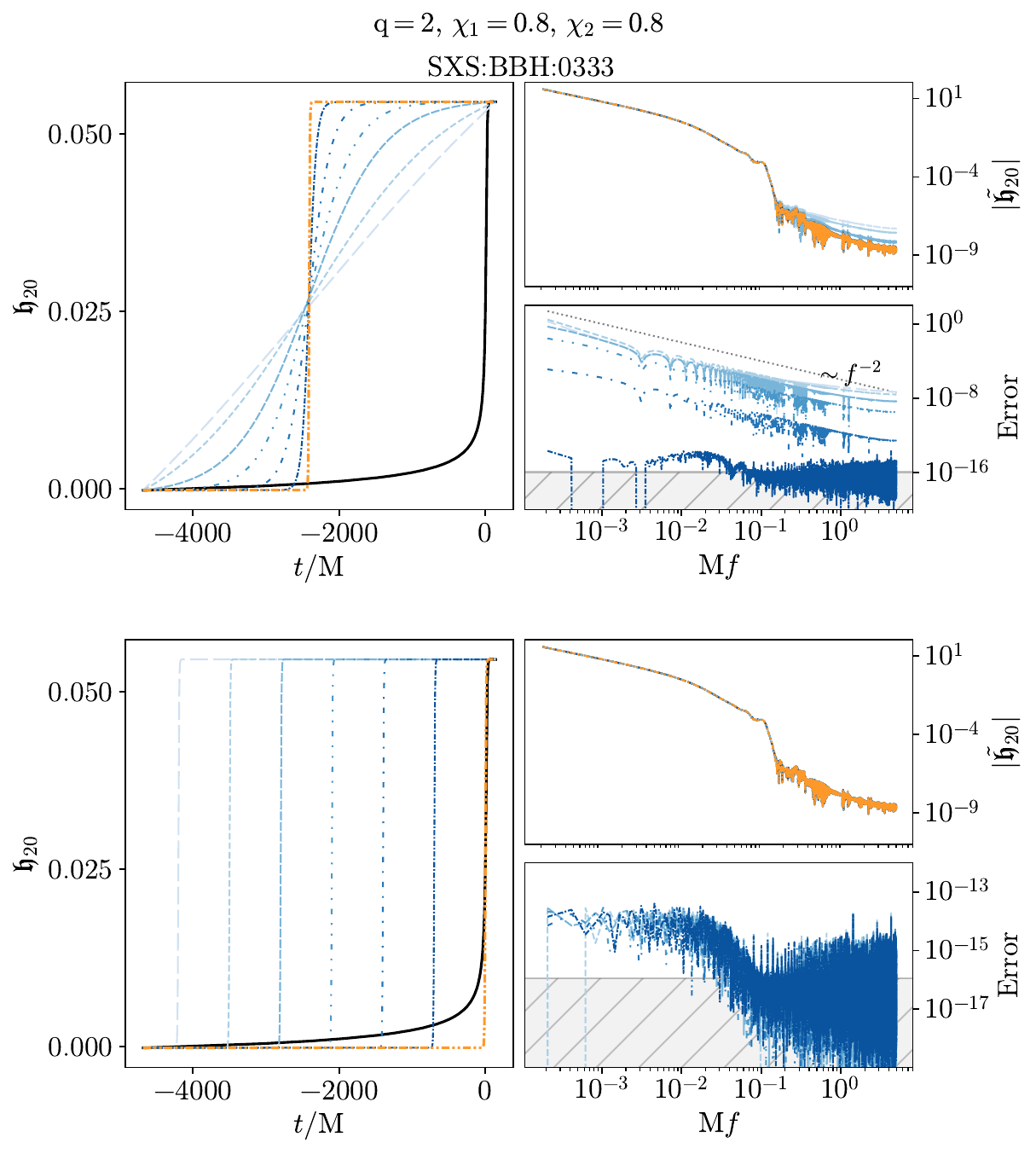}
         \label{fig:y equals x}
     \end{subfigure}
     
     \begin{subfigure}[b]{0.495\textwidth}
         \centering
         \includegraphics[width=\textwidth]{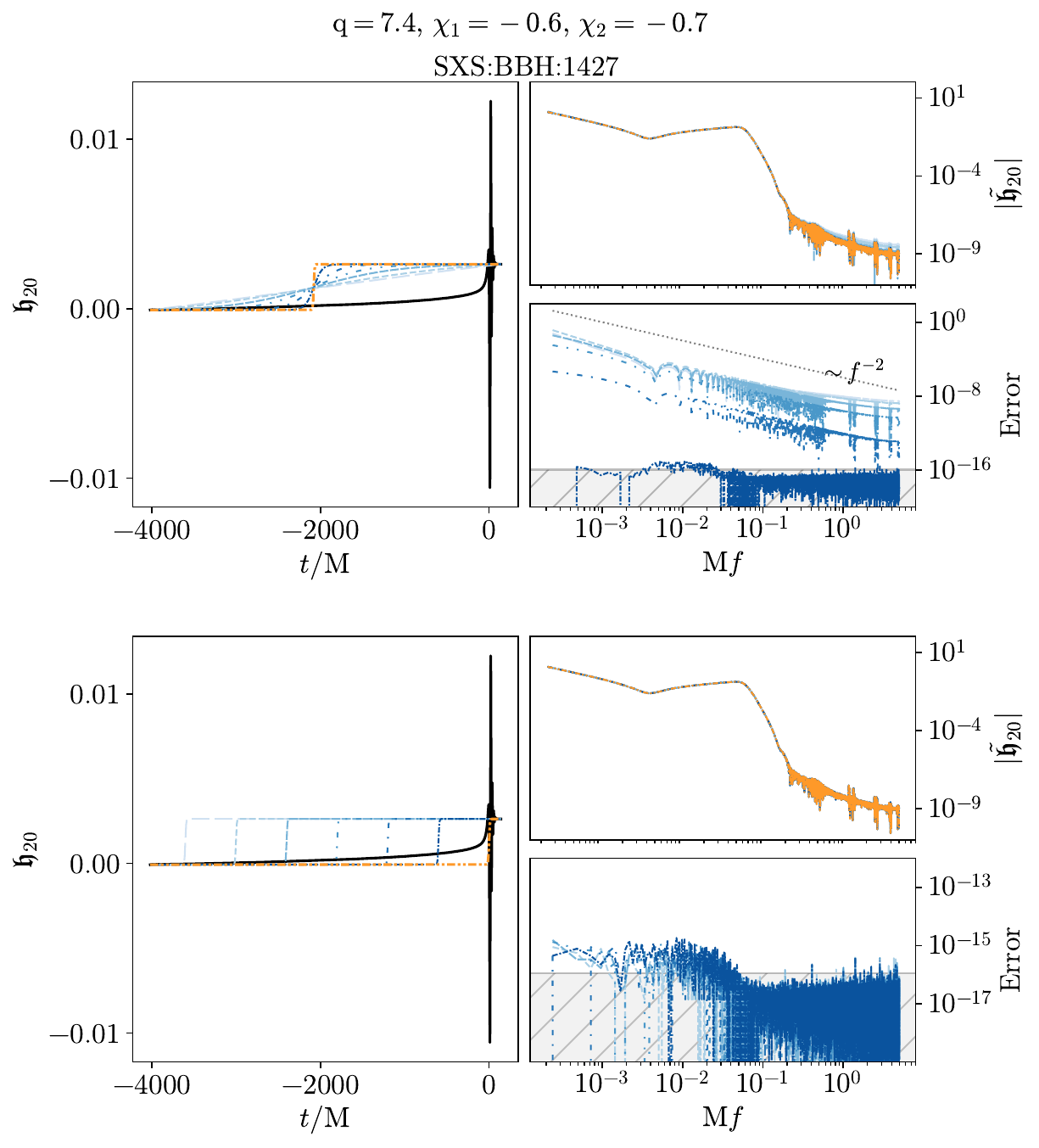}
         \label{fig:three sin x}
     \end{subfigure}
     \hfill
     \begin{subfigure}[b]{0.495\textwidth}
         \centering
         \includegraphics[width=\textwidth]{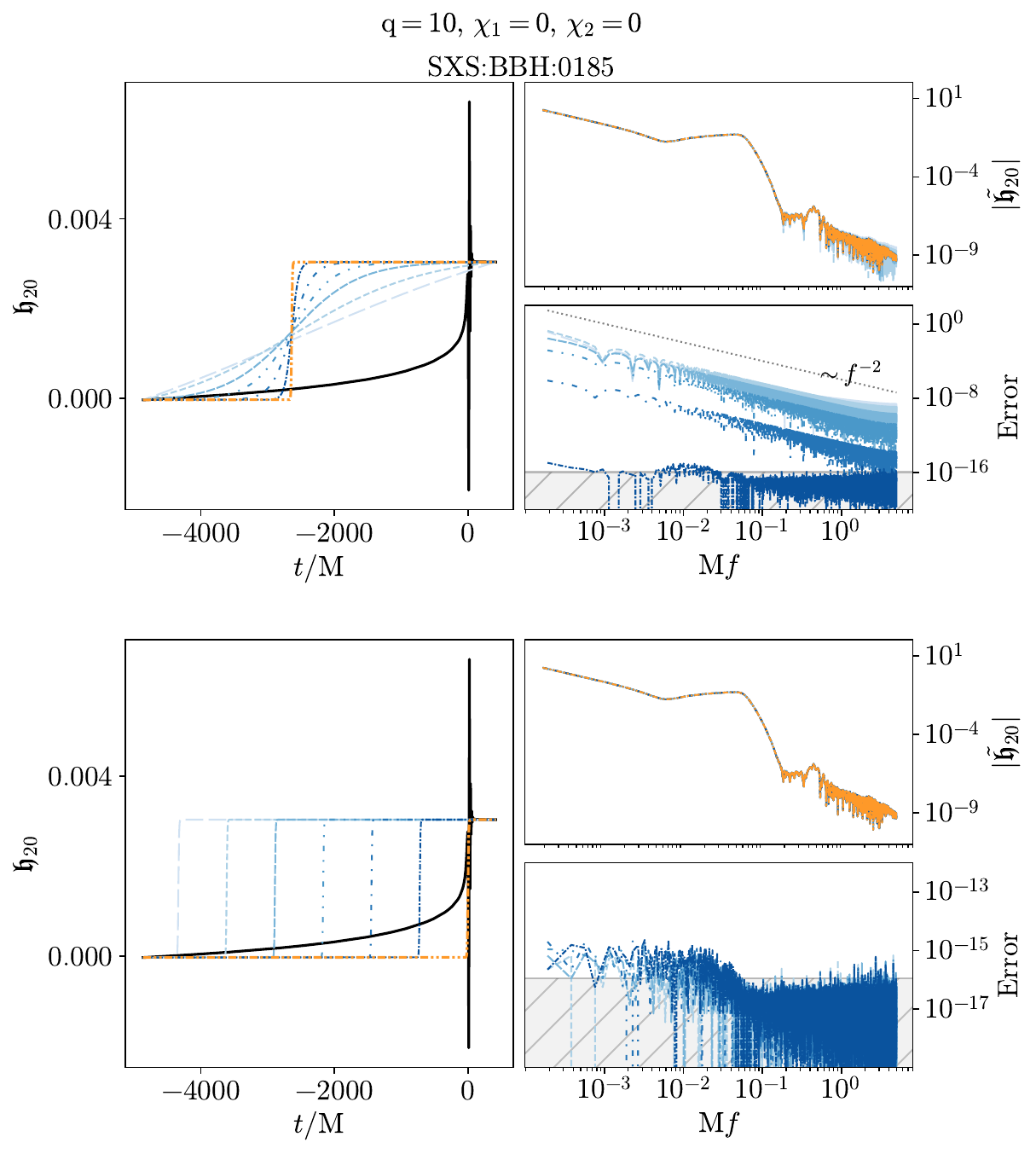}
         \label{fig:five over x}
     \end{subfigure}
     
        \caption{Effect of different choices of $t_{\rm{jump}}$ and $\sigma$ for computing the FT of the (2,0) mode for different NR simulations. For each case, in the upper panels we set $t_{\rm{jump}}$ to be in the middle of the domain and vary \mbox{$\sigma=\{10\,\rm{M},\,100\,\rm{M},\,300\,\rm{M},\,600\,\rm{M},\,1000\,\rm{M},\,2000\,\rm{M},\,5000\,\rm{M} \}$}. In the lower panels we keep $\sigma=10\,\rm{M}$ fix and shift $t_{\rm{jump}}$ along the time axis. The sigmoids used by \texttt{SySS} (blue and orange lines) together with the time-domain (2,0) mode (black line) are displayed in the left panels, whereas the corresponding FTs and the absolute errors with respect to the orange configuration are shown on the right side. The dotted-gray line follow a $\mathcal{O}(f^{-2})$ decay and the gray hatching approximately denotes machine precision.}
        \label{fig:app_C_change_syss_params}
\end{figure*}

In Fig.~\ref{fig:app_C_change_syss_params} we show the impact of the choice of $t_{\rm{jump}}$ and $\sigma$ in \texttt{SySS} for four completely different regions of the parameter space. We have not observed different results in other regions, so have taken these examples as representative cases.

Sigmoids that better capture the boundary behavior of the time-domain signal lead to smoother residuals, 
which will decay faster according to Eq.~\eqref{eq:frequency_decays}. Low values of $\sigma$ are therefore preferred, 
as demonstrated in the upper panels of Fig.~\ref{fig:app_C_change_syss_params}. As commented in Sec.~\ref{subsec:splitting}, excessively low values of $\sigma$, approaching $\sigma\rightarrow 0$, can introduce step-like artifacts in the residual, leading to \textit{aliasing} effects.

For a fix $\sigma$, $t_{\rm{jump}}$ can take any 
arbitrary value along the time axis provided it allows the sigmoid to approach the asymptotic value within the finite time domain, as shown in the lower panels of Fig.~\ref{fig:app_C_change_syss_params}. We take $t_{\rm{jump}}$ at least $3\sigma$ away from the boundaries to achieve accurate results. 

For this paper, choosing $\sigma=10\,\rm{M}$ and $t_{\rm{jump}}=t_{\rm{merger}}$ is enough since the NR waveforms extend beyond 100 $\rm{M}$ after the merger.

Results obtained by \texttt{SySS} show minimal sensitivity to the choice of $t_{\rm{jump}}$ and $\sigma$. They do not need to be tailored to a specific system or region of the parameter space.

\let\c\Originalcdefinition \let\d\Originalddefinition \let\i\Originalidefinition

\bibliography{references}

\end{document}